\documentclass[aps,prd,twocolumn,showpacs]{revtex4}
\usepackage[T2A]{fontenc}
\usepackage[cp1251]{inputenc}
\usepackage[english]{babel}

\usepackage{amsmath}
\usepackage{amsbsy}
\usepackage{amssymb}
\usepackage{graphicx}
\usepackage{calrsfs}

\DeclareMathAlphabet\mathbfcal{OMS}{cmsy}{b}{n}

\newcommand{\aap}{{Astr.~Ap.}}

\newcommand{\sovast}{J. Exp. Theor. Phys.}
\newcommand{\physrep}{Phys. Rep.}
\newcommand{\mnras}{{\em M.N.R.A.S.{\rm}}}

\begin{document}

\def\b{\boldsymbol}
\def\d{\partial}
\def\p{\varpi}
\def\e{\varepsilon}
\def\k{\kappa}
\def\ds{\displaystyle}
\def\t{\tilde}
\def\apjs{ApJS}

\title{On transition of propagation of relativistic particles from the ballistic to the diffusion regime}

\author{A.Yu.~Prosekin}
\affiliation{Max-Planck-Institut f\"ur Kernphysik,
Saupfercheckweg 1, D-69117 Heidelberg, Germany}
\email{Anton.Prosekin@mpi-hd.mpg.de}

\author{S.R.~Kelner}
\affiliation
{Max-Planck-Institut f\"ur Kernphysik,
Saupfercheckweg 1, D-69117 Heidelberg, Germany}
\affiliation{Research Nuclear University (MEPHI), Kashirskoe shosse 31,
115409 Moscow, Russia}

\author{F.A.~Aharonian}
\affiliation{Dublin Institute for Advanced Studies, 31 Fitzwilliam Place,
Dublin 2, Ireland}
\affiliation
{Max-Planck-Institut f\"ur Kernphysik,
Saupfercheckweg 1, D-69117 Heidelberg, Germany}
\email{Felix.Aharonian@mpi-hd.mpg.de}

\date{\today}

\begin{abstract}
A  stationary distribution function that describes the  entire  processes of propagation of relativistic particles, including the transition between the ballistic and diffusion regimes, is obtained. The spacial component of the constructed function satisfies to the first two moments of the Boltzmann equation. The angular part of the distribution provides accurate values for the angular moments derived from the Boltzmann equation, and gives a correct expression in the limit of small-angle approximation. Using the derived function, we studied the  gamma-ray images produced through the $pp$ interaction of relativistic particles with gas clouds in the proximity of the accelerator.  In general, the morphology and the energy spectra of gamma-rays significantly deviate from the ``standard'' results corresponding to the  propagation of relativistic particles strictly 
in the diffusion regime.
\end{abstract}

\maketitle

\section{Introduction}
Propagation of the cosmic rays in the turbulent magnetic fields can proceed in different regimes 
depending on the scales under consideration. On small scales,  when the particles move
coherently, their propagation is ballistic. This usually occurs close to the source, just after the particles escape the sites of their acceleration. With time, the multiple stochastic scattering in turbulent magnetic fields leads to the isotropization of  directions of cosmic rays. The complete isotropization implies that the propagation proceeds in the diffusion regime. 

The limiting cases of small-angle and isotropic particle
distributions allow solutions of  the problem of  particle propagation
on small and large scales, respectively \citep{Dolginov1967,Aharonian2010,Syrovatskii1959}. 
The small-angle approximation fails when the deflection of particles becomes large, typically larger than  one
radian. The solution of the diffusion equation,  in addition to  its inability to  be applied to small spacial scales,  faces the so-called problem of   superluminal propagation \citep{Aloisio2009}.  
In this regard, the apparent  requirement  $r^2/D\gtrsim r/c$ implies that  the diffusion  
works  when  $r\gtrsim D/c$, where $D$ is the  diffusion coefficient.

 In the small-angle approximation, the evolution of  angular distribution has a diffusive behaviour \citep{Dolginov1967,Aharonian2010}. However,  since  the pitch angle changes
within  the limited  interval,   $-1\leq\mu\leq 1$ ($\mu$ is the cosine of the pitch angle), the mean square displacement of the pitch angle deviates from diffusive behaviour as the average deflection angle grows. 
For  the isotropic turbulence,   the moments of the pitch-angle distribution 
on large timescales have an exponential behaviour \cite{Tautz2013}. This means that the isotropization  becomes  fast after the characteristic time  which  is determined by the pitch-angle Fokker-Planck coefficient $D_{\mu\mu}$. On the other hand, a slower isotropization can happen if the turbulent magnetic field has a slab geometry.

In this work  we use the method of moments. It allows us to eliminate the angular dependence at the expense of introduction of  an  isotropization function that determines the dynamics of  isotropization. The final results depend on the form  of the isotropization function. However,  as we demonstrate below,  for a 
 reasonable choice of the form of the function,  the results  remain quite  stable.

To avoid  of the problem of  the superluminal  motion,  Aloisio and Berezinsky  \cite{Aloisio2009}  have 
introduced the so-called J\"{u}ttner function, which describes the evolution of the cosmic ray density. Although this function is obtained phenomenologically from the formal similarity between the diffusion propagator and the Maxwellian distribution, it gives correct results in the limiting cases of diffusion and the ballistic regime. 
Below we will  show that the integrated over time J\"{u}ttner function is close  our stationary solution, which proceeds from Boltzmann equation.

In many cases, especially for the problems  related to  the radiation  of  cosmic rays, it is necessary to know not only  the cosmic-ray density but also their angular distribution. Indeed, the part of  radiation emitted by particles with strongly anisotropic distribution,    can have a strong impact on the morphology and spectrum  of  radiation,
or even simply missed by the observer.

To demonstrate the importance of the angular distribution for calculations of the apparent gamma-ray morphology, we calculated the gamma-ray intensity maps of the regions surrounding the cosmic ray accelerator. The most distinct features can be  seen for the clumps of matter close to the source. The anisotropy changes significantly the radiation  spectrum  and leads  to fast decrease of intensity at high energies. This results in a suppression or a disappearance of radiation from  nearby clouds located away from the line of sight towards to the cosmic-ray source. Moreover,  even in the case of  homogeneous matter distribution,  the effects of anisotropy and the transition from ballistic to diffusion regime,   play an important  role in the formation of gamma-ray morphology.

The article is organised as follows. In Section~\ref{sec:descr}  we describe the formalism based on the Boltzmann equation,  and obtain  the stationary distribution function, which is valid for all scales from the ballistic to diffusion regime.  In Section~\ref{sec:gammorph}, this distribution function is used for  calculations of the gamma-ray morphology of   nearby clouds irradiated by cosmic rays. The conclusions are summarized in Section~\ref{sec:conc}.

\section{Analytical description}\label{sec:descr}
\subsection{Method of moments}
Let us consider the evolution of the distribution function in the case
of multiple stochastic scatterings. Here we do not specify the mechanism of the scattering,  
and  describe the processes only by the generic probability of the particle to be scattered.
The evolution of the distribution function $f(t,\b r, \b n)$ is determined by the Boltzmann 
transport equation
\begin{equation}\label{eq:Boltzmann}
\frac{\d f}{\d t}+\b n \frac{\d f}{\d \b r}=St f+\frac{\delta(t)\delta(\b r)}{4\pi},
\end{equation}  
where the speed of the light (propagation speed of ultrarelativistic particles) 
is set $c=1$. Here $\b n$ is the unit vector in the direction of propagation.
 It is assumed that the particles are produced by an instant spherically symmetric point-like source described by the Dirac delta functions $\delta(t)\delta(\b r)$.
We consider only elastic collisions which are described by the collision integral
\begin{equation}\label{col_integr}
St f=\int f(\b n')w(\b n'\rightarrow \b n)d \Omega'-\int f(\b n)w(\b n\rightarrow \b n')d \Omega',
\end{equation}
where $w(\b n'\rightarrow \b n)$ is the probability  of scattering of the  particle from the initial direction along $\b
n'$ to the final direction along $\b n$ per unit time. In the case of isotropic medium, the probability $w$  depends only on the angle between the initial and
final directions. For compactness of presentation, the dependence of the distribution function on time and coordinates in
Eq.~(\ref{col_integr}) is omitted.

The solution of  Eq.~(\ref{eq:Boltzmann}) can be found in the small-angle approximation which is valid for initial moments of time. Below we will show that  proceeding from this equation one can  obtain also  the 
equation for diffusion of particles, and derive its solution  which is valid for large time intervals.

We are interested, first of all,  in the transition between
these two solutions. For these purpose it is useful to simplify the problem and consider instead of the distribution function its moments. Applying successively the integral operators $\int d\Omega$, $\int d\Omega\, n_{\alpha}$, ... , $\int d\Omega\, n_{\alpha}...n_{\omega}$ to the Boltzmann equation, one can obtain the equations for the moments.
We restrict ourselves to the first two moments: the density $g=\int f d\Omega$ and the current $\b j=\int \b n f d\Omega$. They are  governed by the following equations:
\begin{align}\label{eq:BolMom}
&\frac{\d g}{\d t}+\frac{\d j_{\alpha}}{\d r_{\alpha}}=\delta(t)\delta(\b r),\\\nonumber
&\frac{\d j_{\alpha}}{\d t}+\frac{\d}{\d r_{\beta}}\langle n_{\alpha}
n_{\beta}\rangle g=-\frac{j_{\alpha}}{\tau}.
\end{align}
Here $\tau$ is the scattering time, which is the inverse of the transport cross section $\sigma_{tr}=1/\tau$, where 
\begin{equation}
\sigma_{tr}=\int (1-\b n \b n') w(\b n'\rightarrow \b n)d \Omega.
\end{equation} 
In the derivation of equations in Eq.~(\ref{eq:BolMom}) it has been taken into account that $\int n_{\alpha} d\Omega=0$, $\int St f d\Omega=0$.

The density and the current depend on the higher moments of the distribution function. To close the system of equations given by Eq.~(\ref{eq:BolMom}), 
we should define the form of the isotropization tensor,
\begin{equation}
\langle n_{\alpha}n_{\beta}\rangle=\frac{\int n_{\alpha}n_{\beta} f d\Omega}{\int f d\Omega},
\end{equation} 
based on the following physical arguments. 

In the spherically symmetric case,  the radial direction is the only preferential direction, therefore, the isotropization tensor should have the following structure
\begin{equation}
\langle n_{\alpha}n_{\beta}\rangle=A\delta_{\alpha\beta}+B\rho_{\alpha}\rho_{\beta},
\end{equation} 
where $\b \rho=\b r/r$ is the radial direction. The standard procedure for the determination of the coefficients $A$ and $B$, which consists in the consequent multiplication by tensors $\delta_{\alpha\beta}$ and 
$\rho_{\alpha}\rho_{\beta}$, leads to the equations
\begin{align}
& 1=3A+B\\\nonumber
&\langle (\b n\b\rho)^2\rangle=A+B,
\end{align}
from where  we  find
\begin{equation}
A=\frac{1-\langle (\b n\b\rho)^2\rangle}{2}, \quad B=\frac{3\langle (\b n\b\rho)^2\rangle-1}{2}.
\end{equation}
We assume that the tensor depends only on the coordinates.
Then, it is convenient to introduce the isotropization function $\phi(r)=B$ which would allow us to write 
the isotropization tensor in the form
\begin{equation}\label{eq:isoten}
\langle n_{\alpha}n_{\beta}\rangle=(1-\phi(r))\frac{\delta_{\alpha\beta}}{3}+\phi(r)\rho_{\alpha}\rho_{\beta}.
\end{equation}
The tensor consists of the unidirectional, $\rho_{\alpha}\rho_{\beta}$, and the isotropic, $\delta_{\alpha\beta}/3$, parts. At $r=0$, when $\b n=\b\rho$, we have $\langle n_{\alpha}n_{\beta}\rangle(0)=\rho_{\alpha}\rho_{\beta}$, whereas at the infinity all directions of $\b n$ are distributed isotropically, and $\langle n_{\alpha}n_{\beta}\rangle(\infty)=\delta_{\alpha\beta}/3$. Thus,  the isotropization function should satisfy the boundary conditions
\begin{equation}
\phi(0)=1\quad \text{and}\quad \phi(\infty)=0.
\end{equation}

In the spherically symmetric case, the density and the current are expressed as $g=g(t,r)$ and $\b j=j(t,r)\b\rho$.
Then the substitution of the isotropization tensor in the form given by Eq.~(\ref{eq:isoten}) into Eq.~(\ref{eq:BolMom}) results in
\begin{align}\label{eq:sphsys}
&\frac{\d G}{\d t}+\frac{\d J}{\d r}=\frac{\delta(t)\delta(r)}{4\pi},\qquad \qquad\\\nonumber
&\frac{\d J}{\d t}+\frac{\d G}{\d r}-\frac{1}{r}\frac{\d}{\d r}\left(\frac{2}{3}(1-\phi)rG\right)=-\frac{J}{\tau},
\end{align}
where we have introduced the functions $G=r^2g$ and $J=r^2j$.

At small distances and times, the current changes fast, i.e. $\frac{\d J}{\d t}\gg \frac{J}{\tau}$. Therefore
one can neglect the term in the right-hand side of the second equation in Eq.~(\ref{eq:sphsys}). The condition
$\phi(0)=1$ leads to the cancellation of the third term in the left-hand side of the same equation. Thus, the equations in Eq.~(\ref{eq:sphsys}) are reduced to
\begin{align}\label{eq:sysbal}
&\frac{\d G}{\d t}+\frac{\d J}{\d r}=\frac{\delta(t)\delta(r)}{4\pi},\qquad \qquad\\\nonumber
&\frac{\d J}{\d t}+\frac{\d G}{\d r}=0,
\end{align}
which can be rewritten in the form of the wave equation \mbox{for $G$}
\begin{equation}
\frac{\d^2 G}{\d t^2}=\frac{\d^2 G}{\d r^2}, 
\end{equation}
with the  boundary condition $G(t=+0,r)=\delta(r)/4\pi$. The solution of this equation in terms of the 
density $g=G/r^2$ is
\begin{equation}\label{eq:bal}
g(t,r)=\frac{1}{4\pi r^2}\delta(r-t),
\end{equation}
which describes the behaviour of  density in the ballistic regime.

In the opposite case , i.e. for large distances and times, it is more convenient to proceed from the initial system of equations given by  Eq.~(\ref{eq:BolMom}). The current changes slowly on  scale of the scattering time $\tau$, i.e. 
$\frac{\d j}{\d t}\gg \frac{j}{\tau}$. This allows us to neglect the derivative $\frac{\d j}{\d t}$ in the 
second equation. Taking into account that $\phi=0$ we find
\begin{align}\label{eq:sysdif}
&\frac{\d g}{\d t}=-\nabla\b j,\\\nonumber
&\b j=-\frac{\tau}{3}\nabla g,
\end{align}
which can be rewritten in the form of the  diffusion equation for $g$
\begin{eqnarray}\label{eq:dif}
\frac{\d g}{\d t}=\frac{\tau}{3}\Delta g.
\end{eqnarray}
The comparison with the conventional  form of diffusion equation,  gives the well known
relation
\begin{equation}\label{eq:dtaurel}
D=\frac{c^2\tau}{3}=\frac{cl_c}{3},
\end{equation} 
where $l_c=c\tau$ is the scattering length.

\subsection{Stationary case}
The system of equations derived in the previous section describes the particle motion 
over the entire process of propagation, including the ballistic and diffusion modes, 
as well as the transition stage between these two regimes.  They have  a simple stationary solution. Indeed, the integration over the entire time cancels out the time derivatives and leads to the following system of ordinary differential equations 
\begin{align}\label{eq:sphst}
&\frac{d J}{d r}=\frac{\delta(r)}{4\pi},\\\nonumber
&\frac{d G}{d r}-\frac{1}{r}\frac{d}{d r}\left(\frac{2}{3}(1-\phi)rG\right)=-\frac{J}{\tau},
\end{align}
where $J=J(r)$ and $G=G(r)$.
The first of these equations gives $J=\Theta(r)/4\pi$, where $\Theta(r)$ is the Heaviside step function.
The substitution of the current $J$ to the second equation results in the ordinary differential equation of the 
first order. It's solution can be presented in the form
\begin{equation}\label{eq:stres}
g(r)=\frac{\chi(r)}{4\pi r^2\tau}\int\limits_{r}^{\infty} dr' \exp\left(-\int\limits_{r}^{r'}\psi(r'')\right),
\end{equation}
where
\begin{equation}
\chi(r)=\frac{3}{1+2\phi(r)},\quad \psi(r)=\frac{2}{r}\left(\frac{1-\phi(r)}{1+2\phi(r)}\right).
\end{equation}
The limits for the integrals are chosen from the condition that  the density should vanish in the infinity.

To find the final expression,  one should choose a suitable form of the isotropization function $\phi(r)$. The boundary condition $4\pi g(r)r^2\rightarrow1$ at $r\rightarrow0$, which has not been used yet, gives a  relation between $\phi(r)$ and $\tau$. To take into account this relation, $\phi$ should have one free parameter $\nu$. In the case of isotropic medium, one expects an exponential rate of isotropization \cite{Tautz2013}. Therefore, one of the possible expressions for isotropization function is 
\begin{equation}\label{eq:phie}
\phi(r)=e^{-r/\nu}.
\end{equation} 
However, this function does not allow representation of Eq.~(\ref{eq:stres}) in quadratures. Another expression
\begin{equation}\label{eq:phipl}
\phi=\frac{1}{1+r/\nu}
\end{equation}
is less physically motivated, but allows   us  to obtain a simple analytical solution. Indeed, the substitution of Eq.~(\ref{eq:phipl}) into Eq.~(\ref{eq:stres}) results in
\begin{equation}\label{eq:statsol}
g(r)=\frac{1}{4\pi r^2}\frac{3(\nu+r)}{\tau}.
\end{equation}
The boundary condition  $4\pi g(r)r^2\rightarrow1$ at $r\rightarrow0$ gives $\nu=\tau/3$. The comparison with
Eq.~(\ref{eq:dtaurel})  gives  $\nu=D/c$. Then, the solution can be written in terms of 
diffusion coefficient $D$ as
\begin{equation}\label{eq:dens}
g(r)=\frac{1}{4\pi}\left(\frac{1}{r^2}+\frac{c}{rD}\right),
\end{equation}
which is just the sum of the solutions in the limiting cases of ballistic and diffusion regimes. Introducing a dimensionless parameter $x=rc/D$, which is the radial distance in the units of $D/c$, one can rewrite this expression in the form
\begin{equation}\label{eq:simsol}
g(r)=\frac{(1+x)}{4\pi r^2}.
\end{equation}

\begin{figure}
\begin{center}
\includegraphics[bb=0 0 720 720,width=0.55\textwidth]{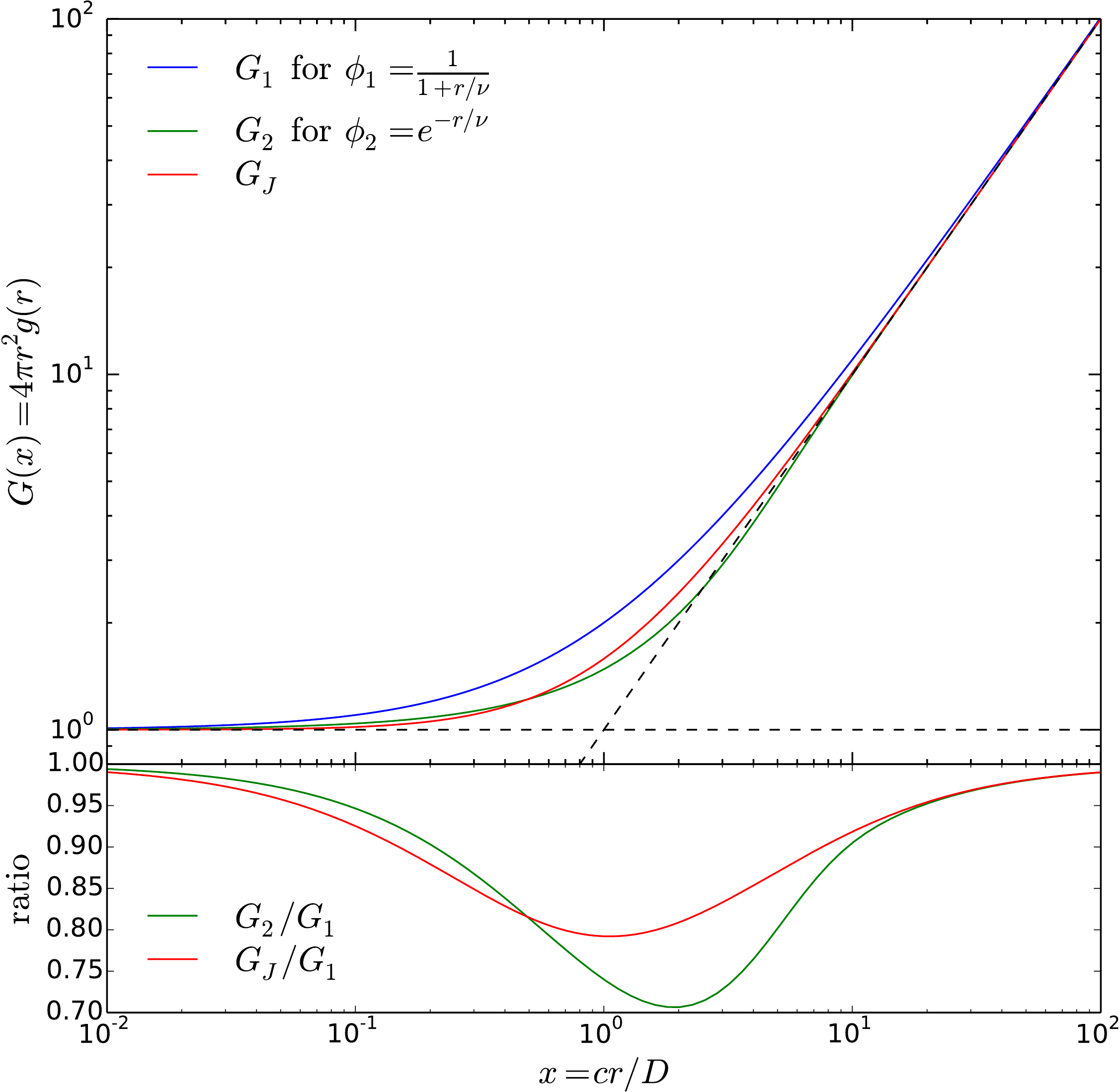}
\caption{\label{fig:Dens} Upper panel: Density $G=4\pi r^2 g(r)$ for different models of isotropization as function of $x=cr/D$. $G_1(x)$ and $G_2(x)$ correspond to 
$\phi_1=\frac{1}{1+r\nu}$ and $\phi_2=e^{-r\nu}$ isotropization functions. $G_J(x)$ corresponds to the integrated over time generalized
J\"{u}ttner function proposed by~\cite{Aloisio2009}. The dashed lines represents the asymptotes $G(x)=1$ and \mbox{$G(x)=x$}. Lower panel: The ratios of the functions: $G_2/G_1$ and $G_j/G_1$.}
\end{center}
\end{figure}

The result is surprisingly simple. However, we should note that it is obtained for a specific form of isotropization function. The results for other forms of isotropization  function can be numerically calculated using Eq.~(\ref{eq:stres}). Fig.~\ref{fig:Dens} shows that  the densities obtained for $\phi=\frac{1}{1+r/\nu}$ and $\phi(r)=e^{-r/\nu}$ differ less than $30\%$, although it should be  noticed  that in the first case the asymptotic behaviour occurs later. For comparison,  we present also the density provided by the generalized
J\"{u}ttner function integrated over time. This function was proposed phenomenologically  
in ref.~\cite{Aloisio2009}  for description of  evolution of the cosmic ray flux. It is seen from Fig.~\ref{fig:Dens} that the J\"{u}ttner function gives a result which is quite close to the solution with an exponential form of the isotropization function.

The propagation of cosmic rays is characterized not only by the density, but also by the angular distribution of particles.  Such an information is contained in the moments of the angular distribution. Indeed, if $\mu=\b n\b \rho$ is the cosine
of the angle between the particle and the radial direction, then the first moment of the angular distribution is
\begin{equation}
\langle \mu \rangle=\frac{\b j\b \rho}{g}.
\end{equation} 
Using the solution given by Eq.~(\ref{eq:simsol}),  we have
\begin{equation}\label{eq:mu1}
\langle \mu \rangle=\frac{1}{1+x}.
\end{equation}
The second moment is just the projection of the isotropization tensor on the radial direction
\begin{equation}\label{eq:mu2}
\langle\mu^2\rangle=\langle n_{\alpha}n_{\beta}\rangle \rho_{\alpha}\rho_{\beta}=\frac{1+2\phi(r)}{3}.
\end{equation}

The moments $\langle \mu \rangle $ and $\langle \mu^2 \rangle $ can be used to construct a function that has properties of the exact angular distribution. In this regard, the simplest function is
\begin{equation}\label{eq:angdis}
M(\mu)=\frac{1}{Z}\exp\left(-\frac{3(1-\mu)}{x}\right),
\end{equation}
where the normalization function has  the form
\begin{equation}
Z(x)=\frac{x}{3}\left(1-e^{-6/x}\right).
\end{equation}
$M(\mu)$ is the distribution of  the cosine of  pitch angles $\mu$ in the range $[-1,1]$. This distribution gives the moments that are in a good agreement, at small and large distances,  with moments given by Eqs.~(\ref{eq:mu1}) and (\ref{eq:mu2}). For both moments the difference from Eqs.~(\ref{eq:mu1}) and (\ref{eq:mu2}) does not exceed $35\%$.

One can show that in the small-angle limit the distribution given by Eq.~(\ref{eq:angdis}) describes the diffusion in angle. Indeed, in this limit the transport cross section becomes
\begin{equation}\label{eq:angdiff}
\sigma_{tr}\approx\int (1-(1-\frac{\theta^2}{2}))
 w(\b n'\rightarrow \b n)d \Omega=\frac{1}{2}\langle\frac{\Delta \theta^2}{\Delta t}\rangle=\frac{D_{\theta}}{2},
\end{equation}
where $D_{\theta}$ is the diffusion coefficient in angular space. Taking into account Eq.~(\ref{eq:dtaurel}) and
the relation $x=rc/D$,  we  obtain
\begin{equation}
M(\mu)\sim\exp\left(-\frac{3D(\theta^2/2)}{rc}\right)=\exp\left(-\frac{\theta^2}{D_{\theta}r/c}\right).
\end{equation}

The combination of the density and the angular distributions results in the stationary distribution function
\begin{equation}\label{eq:dist}
f(r,\mu)=\frac{Q}{4\pi c}\left(\frac{1}{r^2}+\frac{c}{rD}\right)
\frac{1}{2\pi Z}\exp\left(-\frac{3D(1-\mu)}{rc}\right),
\end{equation}
which describes the evolution of propagation,  from ballistic to diffusion, where $Q$ is the source function (the production rate) of cosmic rays in the source.

\subsection{Diffusion coefficient}

In the derived distribution functions  the diffusion coefficient is assumed to be constant in space, but it can have an arbitrary energy dependence.  The latter is determined by  the relation between the  Larmor radius $R_L=E/eB$ and the correlation length of the turbulence $\lambda$. If $R_L\gg\lambda$, the particle is only slightly deflected on the correlation length. The random walk of uncorrelated deflections results in the scattering length $l_c\sim R^2_L$ \citep{Aloisio2004}.  By combining Eq.~(\ref{eq:dtaurel}),  Eq.~(\ref{eq:angdiff}) and the relation $\tau=1/\sigma_{tr}$,  we obtain
\begin{equation}
D=\frac{2}{3}\frac{c^2}{D_{\theta}}.
\end{equation}
The coefficient of diffusion in angle, $D_{\theta}$, can be written as \cite{Aharonian2010}
\begin{equation}
D_{\theta}=\frac{(\alpha-1)(\beta+1)}{4\alpha\beta}\frac{c\lambda}{R_L^2},
\end{equation}
where $\alpha$ and $\beta$ are the power-law indices of turbulence for wave vectors  $k>k_0=2\pi/\lambda$ and $k<k_0$, respectively. The value of $\beta$ is poorly known; here we will assume 
$\beta=1$. 

For the Kolmogorov spectrum of turbulence ($\alpha=5/3$),  calculations of 
the diffusion coefficient in the regime  $R_L\gg\lambda$ give 
\begin{equation}\label{eq:dlarge}
D=\frac{10}{3}c\lambda\left(\frac{R_L}{\lambda}\right)^2 . 
\end{equation}
Note that  the energy dependence in this regime is the same for any other spectrum of turbulence.

At $R_L\ll \lambda$,  particles are only scattered by MHD waves with the length equal to their gyroradius. This leads to the dependence of the  diffusion coefficient   on the turbulence spectrum. The quasi-linear theory predicts $l_c\sim R^{2-\alpha}_L$ \citep{Shalchi2009}, where $\alpha$ is the power-law index of the turbulence spectrum. This dependence can be obtained as follows. The scattering frequency of the particle by 
waves is expressed as \citep{Blandford1987}
\begin{equation}\label{eq:nuscat}
\nu_s=\frac{\pi}{4}\Omega\left(\frac{\mathcal{E}_k k}{B^2/8\pi}\right)_{|k=k_{res}},
\end{equation}
where $\mathcal{E}_k$ is the  spectrum of turbulence normalised  as $\int\mathcal{E}_k dk=B^2/8\pi$,
$k_{res}=\Omega/v\mu$ is the  resonance wave vector, $\Omega$ is the girofrequency, and $\mu$ is the cosine of the pitch angle. The diffusion coefficient is related to the scattering frequency as \cite{Aloisio2004,Shalchi2009}
\begin{equation}\label{eq:dnu}
D=\frac{v^2}{4}\int\limits_{0}^{1}d\mu\frac{1-\mu^2}{\nu_s}.
\end{equation}
Taking turbulence spectrum in the form $\mathcal{E}_k\sim1/k^{\alpha}$ with minimum wave vector $k_0=2\pi/\lambda$, 
and substituting Eq.~(\ref{eq:nuscat}) into Eq.~(\ref{eq:dnu}),  we obtain
\begin{equation}
D=\frac{2}{(2-\alpha)(4-\alpha)(\alpha-1)}\frac{1}{\pi(2\pi)^{\alpha-1}}\lambda c 
\left(\frac{R_L}{\lambda}\right)^{2-\alpha}.
\end{equation}
Thus for $\alpha=5/3$ we have
\begin{equation}\label{eq:dlow}
D=\frac{27}{7}\frac{1}{\pi(2\pi)^{2/3}}\lambda c \left(\frac{R_L}{\lambda}\right)^{1/3}, \quad R_L\ll\lambda.
\end{equation}

We note that  the value of the numerical prefactors  in Eqs.~(\ref{eq:dlarge}) and (\ref{eq:dlow}) 
depend slightly on the  on  turbulence spectrum. The parameter $\lambda$ in these expressions 
corresponds to the largest scale of inertial range of turbulence and can be associated with the 
correlation length.

The simulations of the particle propagation in the isotropic and purely turbulent magnetic field performed 
in  ref.~\cite{Globus2008} show that the diffusion coefficient can be presented in the following form
\begin{equation}\label{eq:diffcoef}
D=\frac{c\lambda}{3}\left(\frac{1}{(2\pi)^{2/3}}\left(\frac{R_L}{\lambda}\right)^{1/3}+\frac{4\pi}{3}\left(\frac{R_L}{\lambda}\right)^2\right).
\end{equation}
In the next section, this convenient expression for the diffusion coefficient 
will be used for calculations of the gamma-ray emission.

\section{Gamma-ray emission of the region surrounding cosmic-ray accelerator}\label{sec:gammorph}
High energy gamma-rays carry unique information not only about the accelerators of cosmic rays (electrons, protons and nuclei), but also allow us to trace these particle after they leave the sites of their acceleration.  In the interstellar medium, this is realized through interactions of cosmic rays with the so-called giant molecular clouds. These dense gas regions illuminated by cosmic rays provide a target for proton-proton interaction and radiate gamma-rays.  Thus they can serve as  unique "barometers" for measurements of the pressure (energy density) of cosmic rays  at different locations relative  to the accelerator.

The massive clouds located in the vicinity of the accelerator dramatically increase the chances of tracing the run-away particles through the secondary gamma-rays. For example, for a young supernova remnant (as an accelerator of cosmic rays) at a distance of $1$ kpc, a gas cloud of mass of order $10^4$ $M_\odot$ can emit very high energy gamma-rays at a level detectable by current instruments if the cloud is located within 100 pc from the supernova remnant \citep{Gabici2009}. Before being fully diffused way and integrated into the "sea" of the galactic cosmic rays,  they  produce gamma-rays the spectrum of which could essentially differ from both the gamma-ray spectrum of the accelerator itself and the spectrum of the diffuse galactic gamma-ray emission. In the case of propagation in a "nominal" diffusion regime, the formation of gamma-ray energy spectra has been discussed in ref.~\cite{Aharonian1996}.  However, closer  to the accelerator, the propagation of cosmic rays may have a more complex character which would be reflected in the spectra of secondary gamma-rays.  

The morphology of the gamma-ray emission is determined by the interplay between cosmic ray and matter distributions. While the distribution of the matter could be arbitrary and random, the density of the cosmic rays decreases with distance from the accelerator, which means the most intensive radiation should arrive from the dense regions located close to the accelerator. 

However, the angular distribution of cosmic rays may lead to a significant deviation from such a simple  picture \citep{Nava2013}. Indeed, due to  the relativistic character of  proton-proton interactions, gamma rays are emitted along the direction of the momentum of the incident proton. It means that only the protons directed towards  the observer give a contribution to the detectable gamma-ray flux. 
Close to the accelerator,  the angular distribution of cosmic rays, which propagate ballistically along the radial direction, is strictly anisotropic. Thus the gamma radiation can arrive from the direction towards accelerator only in the case of presence of  significant amount of gas along the line of sight.  The nearby clouds located not on the line of sight  could be  invisible because their radiation is directed not towards the observer. As the distribution of cosmic rays  becomes more isotropic,  the apparent intensity of the radiation increases with the distance from the accelerator. 

On the other hand, if a gas cloud on the line of sight is located sufficiently close to the accelerator, we could see a bright source which would coincide with the accelerator. This can be misinterpreted as a prolific production of gamma-rays inside the accelerator, although in reality the accelerator could be a very inefficient gamma-ray emitter.

Thus, the consideration of  transition from the ballistic to diffusion regime  may result in two consequences related to the density and angular distributions of cosmic rays. The first one is the  lack of radiation from regions  close to the accelerator  even in the case of presence of massive nearby clouds (but located away from the line of sight).  In contrary, 
we may  detect  very  bright and focused gamma ray  image if  a dense cloud in proximity of the accelerator
would appear  on the line of sight. 

In the following calculations we consider a proton accelerator  of power  
$L_{cr}=10^{37}\, \text{erg}/\text{s}$  located  at a distance $d=1$ kpc from the observer. The energy spectrum of protons is taken in the form $J_{cr}(E)=E^{-2}\exp\left(-E/10^{15}\, \text{eV}\right)$. The stationary distribution of cosmic rays around the accelerator is described by Eq.~(\ref{eq:dist}) with the diffusion coefficient given by Eq.~(\ref{eq:diffcoef}). We assume that particles propagate through a turbulent magnetic field with the Kolmogorov spectrum of turbulence.  We keep the coherence length of 
turbulence $\lambda$ as a free parameter.

In calculations, the magnetic field is taken at the level of $B=10^{-4}$ G. In the galactic disk,  the typical value of the diffusion coefficient is \mbox{$D\approx 10^{28} \text{cm}^2/\text{s}$} at $1$ GeV \citep{Ptuskin2006}. To match this value we assume $\lambda=10^4$ pc and \mbox{$\lambda=10^5$ pc} for the diffusion coefficients, which below we refer to as small and large, respectively. The smallness of the ratio $R_L/\lambda\approx 1.1\times 10^{-9} E_{12}\lambda_{4}^{-1}$, where $E_{12}=E/10^{12}\,\text{eV}$ is the cosmic ray energy and \mbox{$\lambda_{4}=\lambda/10^4\,\text{pc}$}, allows us to neglect the high-energy non-resonant part of the diffusion coefficient. The estimate based on Eq.~(\ref{eq:diffcoef}) gives \mbox{$D\approx 9\times 10^{28} E_{12}^{1/3}\lambda_{4}^{2/3}\,\text{cm}^2/\text{s}$}.
Correspondingly, the transition from ballistic to diffusion regime occurs at a characteristic distance \mbox{$D/c\approx 1.0\, E_{12}^{1/3}\lambda_4^{2/3}\,\text{pc}$} from the cosmic-ray source.

Below we consider  two cases: (i) a homogeneous cloud surrounding the accelerator, (ii)  a group of clouds located near the accelerator. We consider  different  densities  of the background gas (between the clouds),  and different  values of the diffusion coefficient of cosmic rays. For calculation of gamma-ray production in $pp$ collisions the parametrisation of ref.~\cite{Kafexhiu2014} has been used. The results present the intensity of  gamma-radiation integrated along the line of sight through the gamma-ray production region.

In the first example, we consider a   homogeneous cloud with a radius $R=10$ pc and density $n_p=100\,\text{cm}^{-3}$. At these parameters the mass of the cloud is $M_{cl}\approx 10^{4}\, M_{\odot}$.
 The accelerator is placed in the centre of the cloud. The intensity maps of  gamma rays at three different energies, $E=10^{10}$ eV, $E=10^{11}$ eV, and $E=10^{12}$ eV,  are shown in Fig.~\ref{fig:IntHom}. 
Calculations are performed for  the case of  slow diffusion. It is assumed that the density inside the accelerator is very small, thus the gamma-radiation of the accelerator itself can be ignored. 

Fig.~\ref{fig:IntHom} shows that with the increase of energy the radiation becomes more concentrated on  the position of the accelerator.  The change of the gamma-ray morphology is explained by the energy dependence of the cosmic-ray propagation. With the increase of energy, the transition from ballistic to diffusion regime occurs at larger distances from the accelerator. The faster decrease of the cosmic-ray density in the ballistic regime results in the faster decrease of gamma-ray intensity at high energies. It is seen in Fig.~\ref{fig:HmPd} in which the radial profiles of intensity at different energies are shown. The left and right panels of Fig.~\ref{fig:HmPd} present the results calculated for the  small and  large  diffusion coefficients, respectively. For the fast diffusion, the transition from the steep to the flat part of the profile occurs at large distances from the accelerator. 

In Fig.~\ref{fig:HmPd} we show the radial profiles of gamma-ray intensities for  two different regimes of diffusion. To describe the behaviour of these curves, we also show the local power-law indices $\alpha=d\ln P(r)/d\ln(r)$ at different energies. One can see that at small distances
all profiles approach to the same inclination with $\alpha\approx -1.3$. At small energies the flat part starts earlier and intensity decreases slower. For example, at energy $E=10^9$ eV the flattest part of the profile corresponds to $\alpha\approx -0.3$. With an increase of energy, the flat part becomes steeper. At very high energies the transition to the flat part might not happen at all.

The angular distribution of cosmic rays has a strong impact on the characteristics of the secondary 
gamma-radiation, in particular on the radial profile of the gamma-ray intensity 
(see  Appendix~\ref{sec:intpr}).   In the case of strictly radial distribution of particles (i.e. for the 
the ballistic regime 
of propagation),   the gamma-ray source will be  detected  as a point like object, independent of the 
linear size of the gamma-ray production region, L.  For the ``nominal''  diffusion regime,  the angular size 
of the gamma-ray source  is determined by the ratio $L/R$.   The transition regime introduces non-neglible corrections to the formation of the overall image of the gamma-ray source, therefore should be treated thoroughly.   The comparison of  the results in Fig.~\ref{fig:HmPd} 
and Fig.~\ref{fig:SphProf} (which does not take into account the 
angular distribution  of protons)  shows that,    
in general, the shapes of the radial profiles  in these figures are similar, but  there is 
also a significant  difference. In particular, if one assumes that angular distribution of particles is isotropic
just after  their escape from the source,the power-law index of the slope of the 
(projected) gamma-ray profile would be  $\alpha=-1$. This is in contrast to $\alpha\approx -1.3$, which  
is  expected if we correctly treat the angular distribution of particles  closer to the source.  The sharp  angular distribution of cosmic rays close to the accelerator leads to considerable loss of emission, and therefore to a steeper gamma-ray intensity profile. 

The spectral energy distributions (SED) of gamma rays at different distances from the source of cosmic rays are shown in Fig.~\ref{fig:HmS}. The left and right panels show the results corresponding to the small and large diffusion coefficients, respectively. It is seen that the increase of the diffusion coefficient leads to harder gamma-ray spectra.
In the case of homogeneous distribution of the cosmic-ray density and for the power-law energy spectrum of protons  with an index $\alpha=2$,   gamma-rays have an almost flat SED. The flat part slightly  deviates from the cosmic ray spectrum, namely it contains an intrinsic hardening due to the increase of inelastic cross section  \cite{Kelner2006}. For  the  homogeneous distribution of cosmic rays, the flat part of the gamma-ray spectrum can be approximated by power-law with an photon index $\Gamma=1.94$. The SED in the direction to accelerator becomes even harder with the power-law photon index $\Gamma=1.86$. This can be explained by the fact that high-energy protons preserve longer their radial direction and, consequently, produce higher energy radiation towards observer. 

Because of diffusion of cosmic rays, in the directions far from direction to the accelerator, the density of the low-energy protons decreases slower.  At large distances from the accelerator, the SED of gamma-rays is close to  $E^{-\delta}$ (slightly harder because of the  $pp$  interaction cross-section), where  $\delta$ characterizes the energy-dependence of the diffusion coefficient, $D(E)=D_0 E^\delta$. Because of the assumed dependence of the diffusion coefficient with 
$\delta=1/3$, the SED at large distances in Fig.~\ref{fig:HmS} follows $\propto  E^{-1/3}$ behaviour.

In more realistic scenarios,  the surroundings of the cosmic-ray source could be inhomogeneous, i.e. 
may consist of clumps of matter. To study  the general features of radiation of such an environment,  
we use a  simplified  gas distribution-template   consisting of  four identical equally 
separated clouds surrounded by a homogeneous low-density gas (background):
\begin{equation}
n_p=n_{p0}\left(\sum_{i}e^{-\left(\frac{\b r-\b r_{i}}{w_i}\right)^2}+x_{bg}\right),
\end{equation}
where $n_{p0}=10^3\, \text{cm}^{-3}$, $\b r_i$ and $w_i$ are the coordinates of the centres of these clouds and their widths, respectively; $x_{bg}$ is the level of the background relative to the maximum density in the centres of the clouds. The width of each cloud is $w_i=1$ pc which corresponds to the mass $M_{cl}\approx 140\, M_{\odot}$. Separation between clouds is $5$ pc. We consider the case without background and with the background of the level of $x_{bg}=10^{-2}$. The zero level of the background  allows us to eliminate the radiation from the direction towards the accelerator.  For illustrative purposes,  the source of cosmic rays is located  at  the left border of each map in Fig.~\ref{fig:CbLdIm} and Fig.~\ref{fig:CbHdIm}. It is assumed that the  density inside the accelerator is very low, thus its own gamma-radiation can be neglected.

The intensity maps for the case without background gas calculated for the slow and fast diffusion of cosmic rays,
are shown in Figs.~\ref{fig:CwLdIm} and \ref{fig:CwHdIm}, respectively. The corresponding intensity profiles are presented in the left and right panels of Fig.~\ref{fig:CwP}. For slow diffusion, the  gamma-ray intensity of clouds decreases with the distance from the proton accelerator. The clouds are located along the line perpendicular to the line of sight. At high energies in the case of fast diffusion a part of the radiation from the closest clouds is "lost". Therefore, one can see an interesting effect when  despite the  decrease of the cosmic ray density with distance, highest energy gamma-rays  are seen from the furthest rather than closest clouds.

The results corresponding to the  homogeneous gas background are shown in  Figs.~\ref{fig:CbLdIm} and \ref{fig:CbHdIm}. The intensity profiles shown in Fig.~\ref{fig:CbP} are smoother compared to the relevant curves in the case of background absence. The radiation towards the accelerator  appears in Fig.~\ref{fig:CbP} because of the presence of the background gas  on  the  line of sight.  The results in Fig.~\ref{fig:CbP} show  that  in the case of  fast diffusion the gamma-ray intensity is reduced; at very high energies it disappears at all from  the closest cloud.

\begin{table}
\begin{center}
    \caption{\label{tab:CbH}Power-law index~$\alpha$ for the fit to the gamma-ray intensity profiles  for positions of the  maximum radiation (the centres of the clouds) shown in the right panel of Fig.~\ref{fig:CbP} (large diffusion coefficient).}
   \begin{tabular}{ | c | c | c | c | c | c | c |}
       \hline
       $E_{\gamma}$, eV & $10^{9}$ & $10^{10}$ & $10^{11}$ & $10^{12}$ & $10^{13}$ & $10^{14}$\\ \hline
       $\alpha$ & -0.85 & -0.73 & -0.49 & -0.21 & 0.53 & 1.14\\ \hline
   \end{tabular}
\end{center}
\end{table}

To describe quantitatively how fast the gamma-ray intensity decreases with distance, the intensity at the position of the maximum (in the centres of the clouds) shown in Fig.~\ref{fig:CbP} has been fitted with power law. The calculated power-law indices of the fits for different energies are presented in Table~\ref{tab:CbH}. It is seen that with increase of energy the profile becomes flatter. The intensity decreases slower in the case of large diffusion coefficient. Moreover, for energies $E=10^{13}$ and $E=10^{14}$ the intensity increases with distance for the farthest two clouds. In all cases the intensity in maximum points changes with distance slower than $1/\rho$, where $\rho$ is the projected distance from the source.

The spectral energy distributions of gamma rays in the direction to the centres of the clouds are shown
in Figs.~\ref{fig:CbS} and \ref{fig:CwS}. The clouds are numbered in the order of their distances  to the accelerator.   The energy spectra of gamma-rays from the clouds are steeper than in the case 
of the homogeneous cloud surrounding the accelerator (see Fig.~\ref{fig:HmS}). Obviously, this is explained by anisotropic distribution of cosmic rays closer to the accelerator.  The picture becomes smoother  if the clouds are embedded in a homogeneous  background gas.

\section{Conclusion}\label{sec:conc}
One of the major issues in the general problem of identification of sources of galactic and extragalactic cosmic rays, is the character of their propagation through the turbulent magnetic fields  outside the accelerators.  Depending on the distance to the source, the level of the magnetic turbulence of the ambient medium, as well as  the energy of particles, their propagation can proceed in the ballistic or in the diffusion regimes.  While the diffusion of cosmic rays  has been comprehensively studies in the literature, the description of propagation in the intermediate  stage, i.e. at the transition from the ballistic to the diffusive  regime,  is a problem of greater complexity regarding  the  exact analytical solutions. To study the dynamics of this transition,  we used a simplified approach. Based on the treatment of  moments of the Boltzmann equation, we derived the system of equations  for the  time evolution of the distribution function of cosmic rays. 
Written in the form of Eq.~(\ref{eq:sphsys}), it  describes  not only the ballistic and diffusion regimes of propagation as limiting cases, but also the transition stage between these  two regimes. 
This  system of equations allows a simple stationary solution  in the form of  Eq.~(\ref{eq:stres}).

The  key feature  of this approach is the proper choice of a specific function which describes the pitch-angle isotropization. Within the chosen approach,  the isotropization  function cannot be strictly determined.  Nevertheless it can be chosen and introduced in a self-consistent manner based on reasonable physics arguments. In this paper, two forms of the isotropization function has been considered: (1) $\phi=e^{-r/\nu}$ and (2) $\phi=1/(1+r/\nu)$. The exponential form seems to be better justified  given that at the presence of  isotropic turbulence  the pitch angle distribution moments behave exponentially~\cite{Tautz2013}. On the other hand the form $\phi=1/(1+r/\nu)$ provides a simple representation of the stationary solution given by Eq.~(\ref{eq:statsol}). The results obtained for both forms of the isotropization function are in agreement within better than $30\%$ accuracy. They agree well also with the integrated J\"{u}ttner function  proposed in ref.~\cite{Aloisio2009}.

The angular, energy and radial distributions of cosmic rays outside the accelerator, in the general case of their propagation, including the transition stage between the ballistic and diffusion regimes, is described by a surprisingly simple function given by Eq.~(\ref{eq:dist}). 

The anisotropy of angular distribution of cosmic rays before they enter the diffusion regime of propagation, leads to a partial or complete "loss"  of  secondary gamma-rays by the observer. The numerical calculations of secondary gamma-rays performed for both the homogeneous and clumpy (consisting of dense clouds) distributions of gas in the vicinity of the accelerator, demonstrate the strong impact of  the effects of particle propagation in the pre-diffusion regime on the apparent gamma-ray morphology and the energy spectrum. Therefore, the detailed studies of spectral and morphological features of high energy gamma-ray emission outside  the  detected (or potential) cosmic ray accelerators can tell us about the propagation character of cosmic rays after they leave the sites of their acceleration.

\appendix
\section{Intensity profiles}\label{sec:intpr}
\renewcommand\thefigure{\thesection.\arabic{figure}}
\setcounter{figure}{0}
\begin{figure*}[t]
\centering
\begin{minipage}[t]{0.45\textwidth}
\includegraphics[bb=0 0 650 632,width=\textwidth]{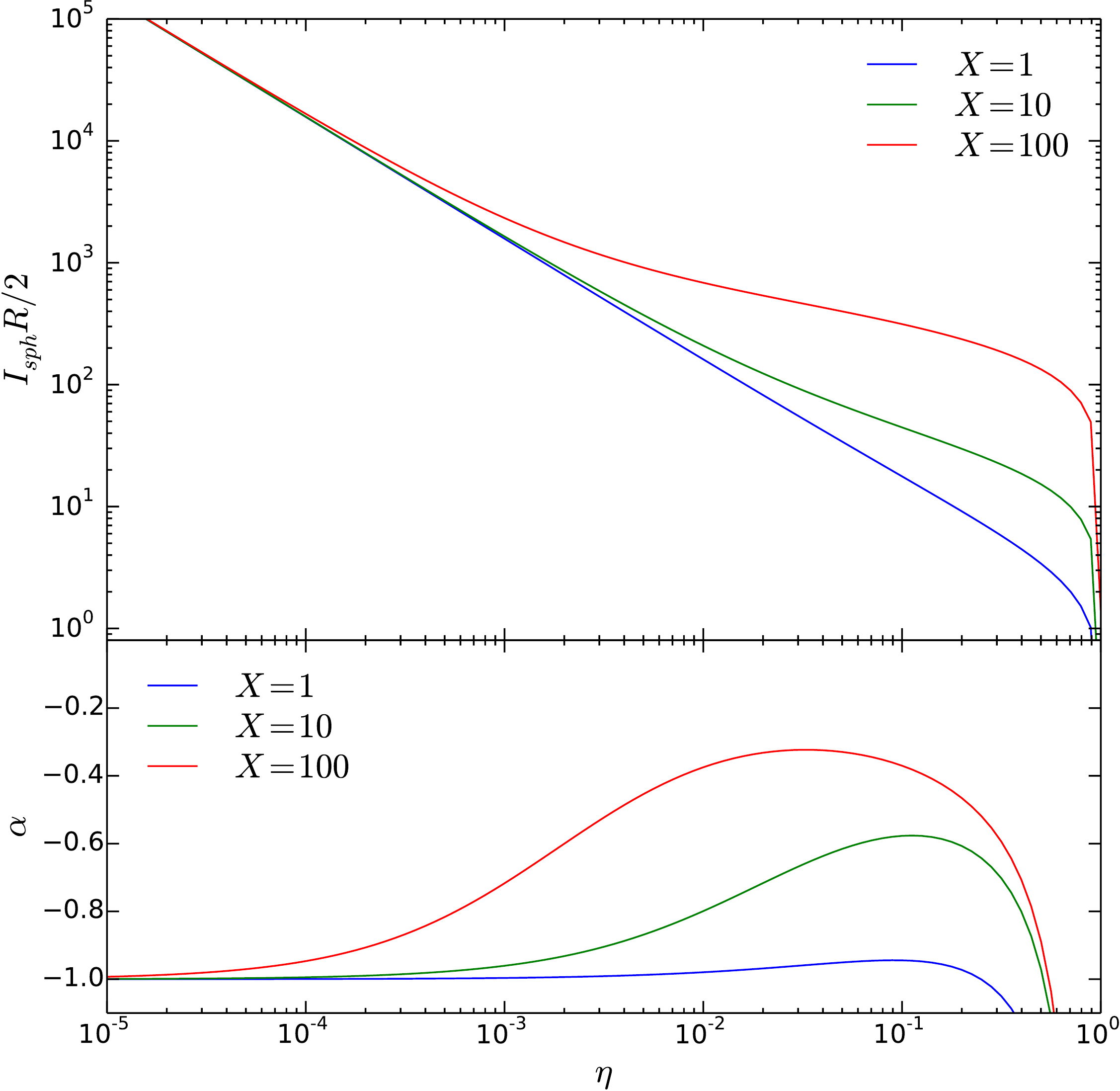}
\caption{\label{fig:SphProf} The gamma-ray intensity profiles calculated for a spherical  homogeneous cloud surrounding the source of protons. The curves correspond to three values of the parameter $X$. The upper panel: the intensity, lower panel - the slop of the profile.}
\end{minipage}
\hspace{.05\linewidth}
\begin{minipage}[t]{0.45\textwidth}
\includegraphics[bb=0 0 650 632,width=\textwidth]{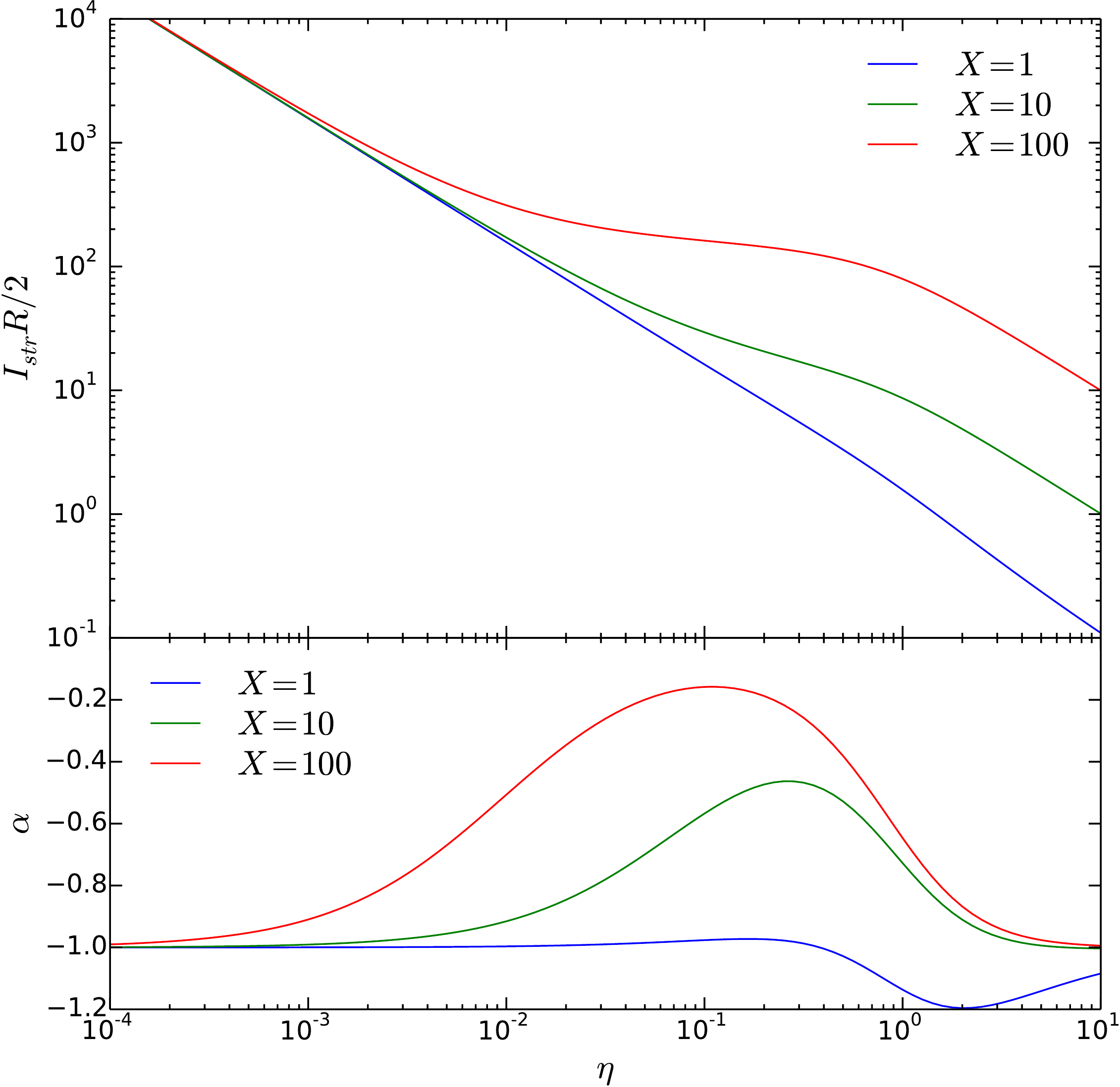}
\caption{\label{fig:StrProf} The same as in Fig.~\ref{fig:SphProf} but for a strip region.}
\end{minipage}
\end{figure*}

The gamma-ray intensity profile corresponds to the radiation integrated along the line of sight and considered as a function of projected distance from the source. Let us assume that the angular distribution of cosmic-ray protons is isotropic. In this case the production rate of gamma rays in  the direction of the observer is proportional to the density of protons. Since the cosmic-ray density changes according to Eq.~(\ref{eq:dens}), the form of intensity profile is
\begin{equation}
I=\int\limits_{z_1}^{z_2} \left(\frac{1}{r^2}+\frac{c}{rD}\right) dz,
\end{equation} 
where $z$ axis is directed along the line of sight.

If the region emitting gamma-rays is a spherical cloud of radius $R$ with the centre at  the position of the particle accelerator, then the integration limits $z_{1,2}=\pm\sqrt{R^2-\rho^2}$ give
\begin{equation}
I_{sph}=\frac{2}{R}\left(\frac{\arctan\sqrt{\frac{1}{\eta^2}-1}}{\eta}+X\mathrm{arccosh}\left(\frac{1}{\eta}\right)\right),
\end{equation}
where $\rho$ is the projected distance from the source, $\eta=\rho/R$, and $X=Rc/D$.
If the emitting region is a strip with width $2R$, and the source of cosmic rays is in the middle of the strip, then the integration limits $z_{1,2}=\pm R$ give
\begin{equation}
I_{str}=\frac{2}{R}\left(\frac{\arctan\frac{1}{\eta}}{\eta}+X\mathrm{arcsinh}\left(\frac{1}{\eta}\right)\right).
\end{equation} 

The functions $I_{sph}R/2$ and $I_{str}R/2$ are presented in Fig.~\ref{fig:SphProf} and Fig.~\ref{fig:StrProf}, respectively. The lower panels show the local slops of
the curves defined as $\alpha=d\ln F(\eta)/d\ln(\eta)$, where $F(\eta)=IR/2$.

It is seen from Figs.~\ref{fig:SphProf} and \ref{fig:StrProf} that  at small distances the intensity changes as $1/\rho$. For $X=1$ this behaviour approximately retains at all distances. This curve corresponds to the case when the diffusion regime has not been reached at the distance $R$. Other curves reveal the transition to the diffusion regime becoming flatter with increase of distance. The integration over smaller region along the line of sight in the case of a spherical cloud makes the slope of the profile steeper compared to the case of strip region.

\renewcommand\thefigure{\arabic{figure}}
\setcounter{figure}{1}

\begin{figure*}
\begin{center}
\includegraphics[bb=0 0 1440 504,width=1\textwidth]{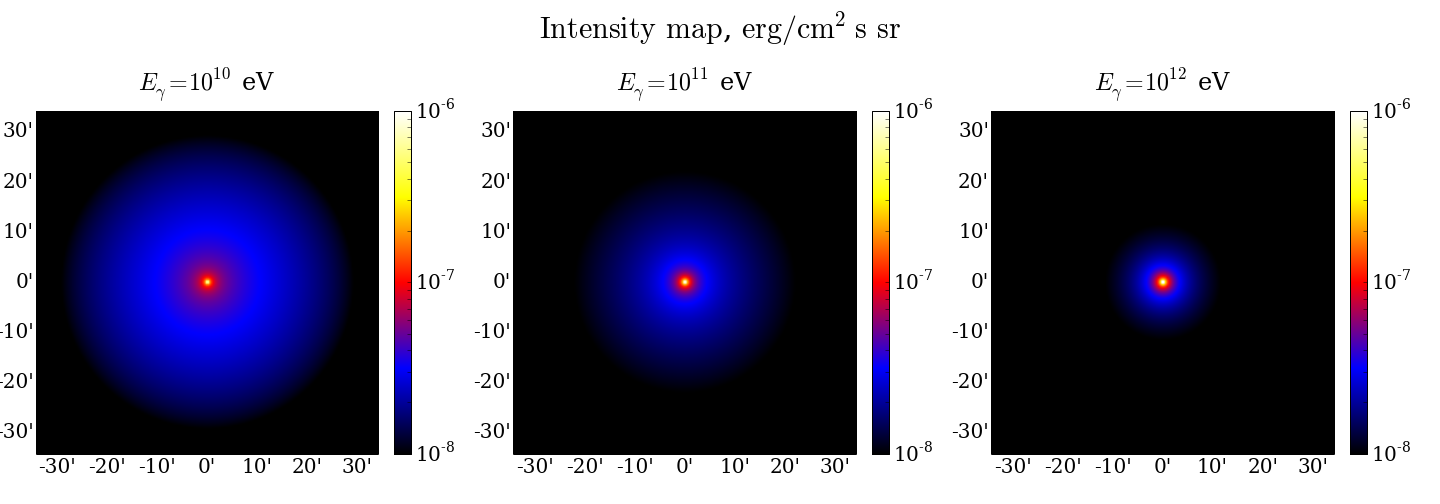}
\caption{\label{fig:IntHom} The intensity maps of gamma-ray emission  at different energies. The spherical cloud with homogeneous density distribution is irradiated by the cosmic-ray source located in its centre.  The  gas density inside the accelerator is assumed very low, so the contribution of the accelerator to the gamma-ray emission is negligible. 
The maps are  produced  for the case of  small  diffusion coefficient (for details, see the text). For the distance to the source  $d=1$ kpc, the region  of $\sim1^{\circ}\times1^{\circ}$ corresponds to the area $\sim20\times20 \ \rm pc^2$.}
\end{center}
\end{figure*}

\begin{figure*}
\begin{center}
\mbox{\includegraphics[bb=0 0 674 646,width=0.5\textwidth]{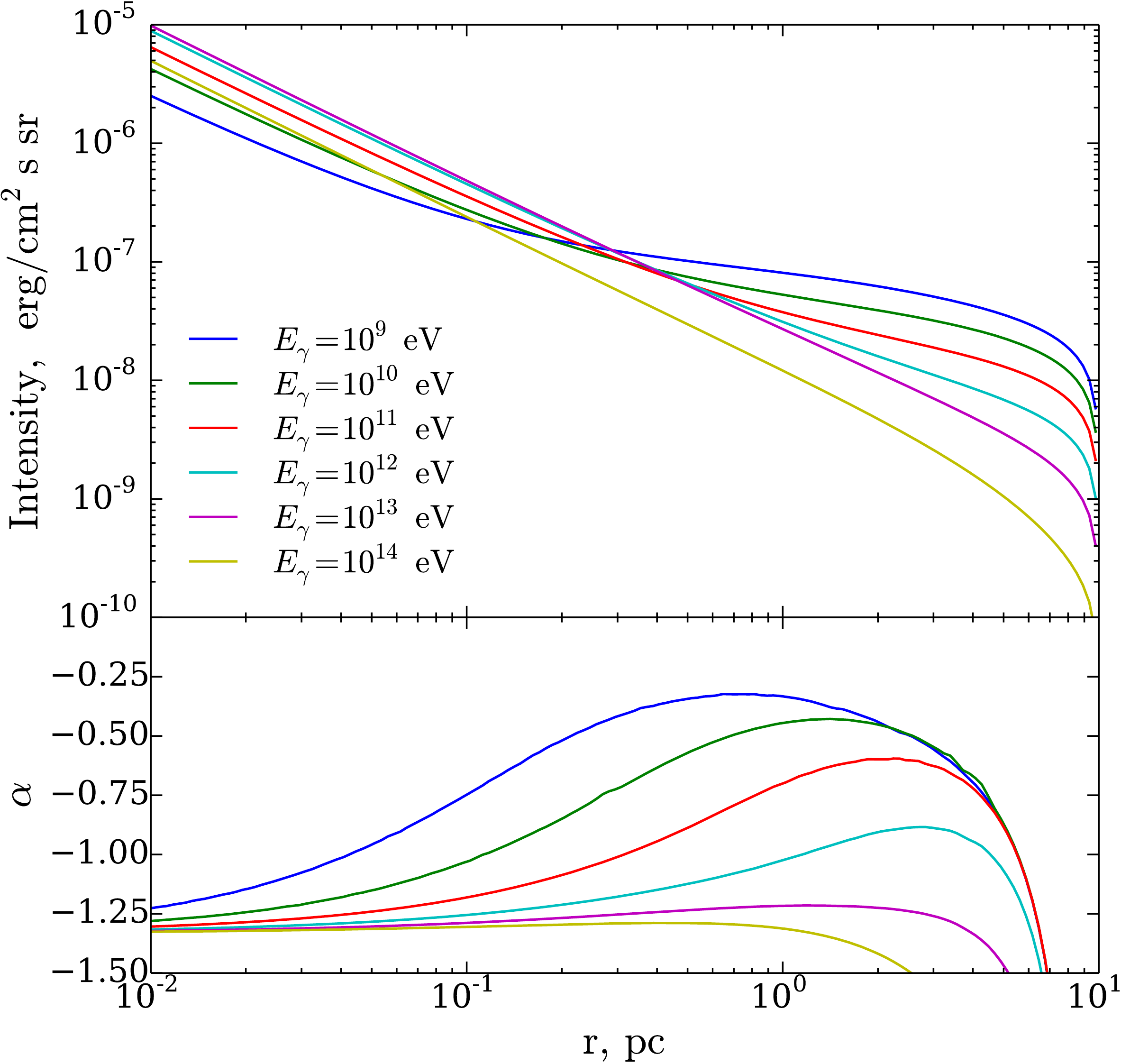}
\includegraphics[bb=0 0 673 646,width=0.5\textwidth]{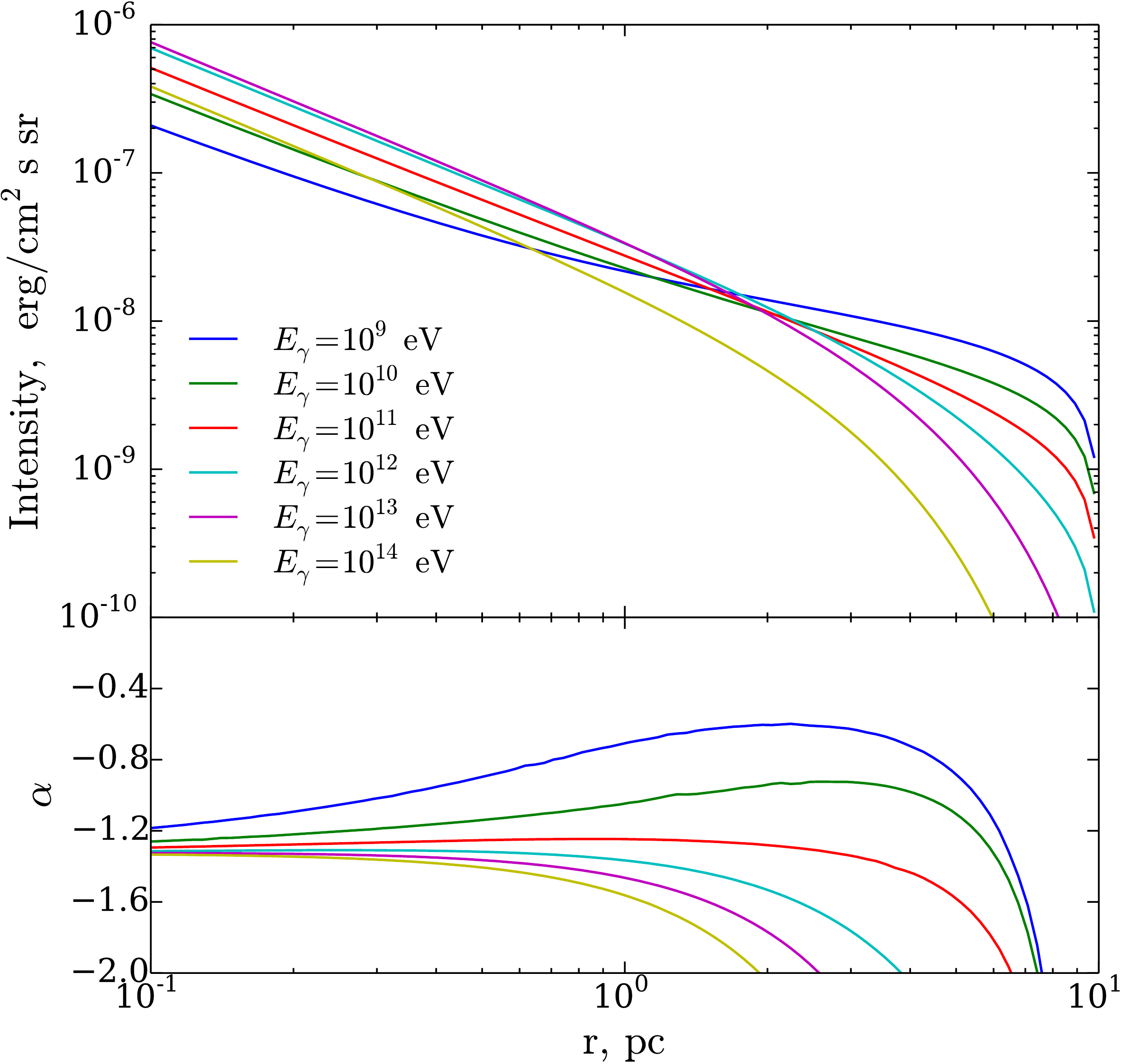}}
\caption{\label{fig:HmPd} Upper panel:  radial intensity profiles for  different energies of gamma-rays in the case of homogeneous cloud surrounding the cloud.  Lower panel: the power-law index of the corresponding intensity profiles in the upper panel. The left and right panels  present the  results calculated for the case of small and large diffusion coefficients, respectively.
The profiles for energies $E=10^{10}$ eV, $E=10^{11}$ eV, and $E=10^{12}$ eV in the left panel correspond to the intensity maps  shown  in Fig.~\ref{fig:IntHom}.}
\end{center}
\end{figure*}

\begin{figure*}
\begin{center}
\mbox{\includegraphics[bb=0 0 678 472,width=0.5\textwidth]{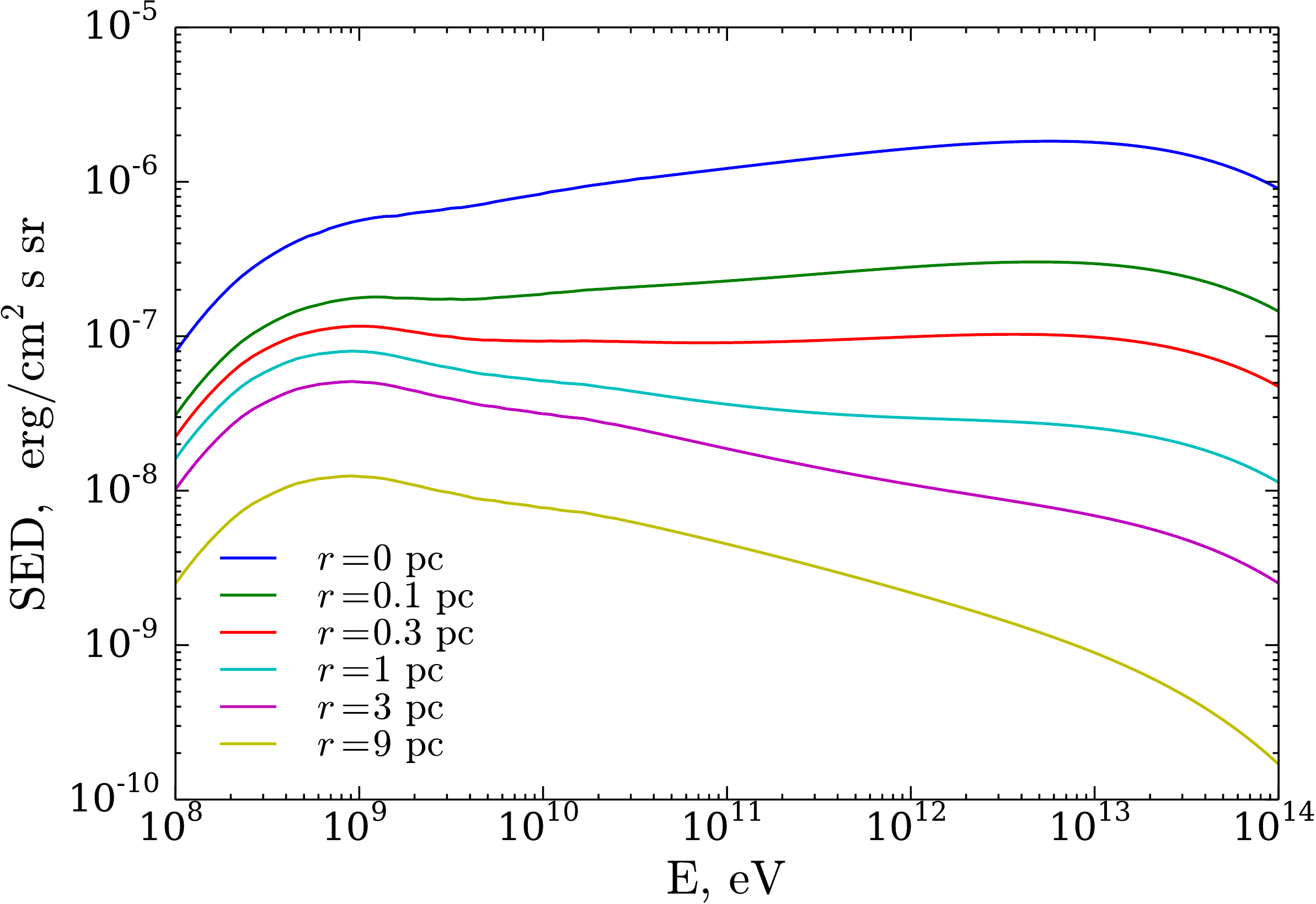}
\includegraphics[bb=0 0 678 472,width=0.5\textwidth]{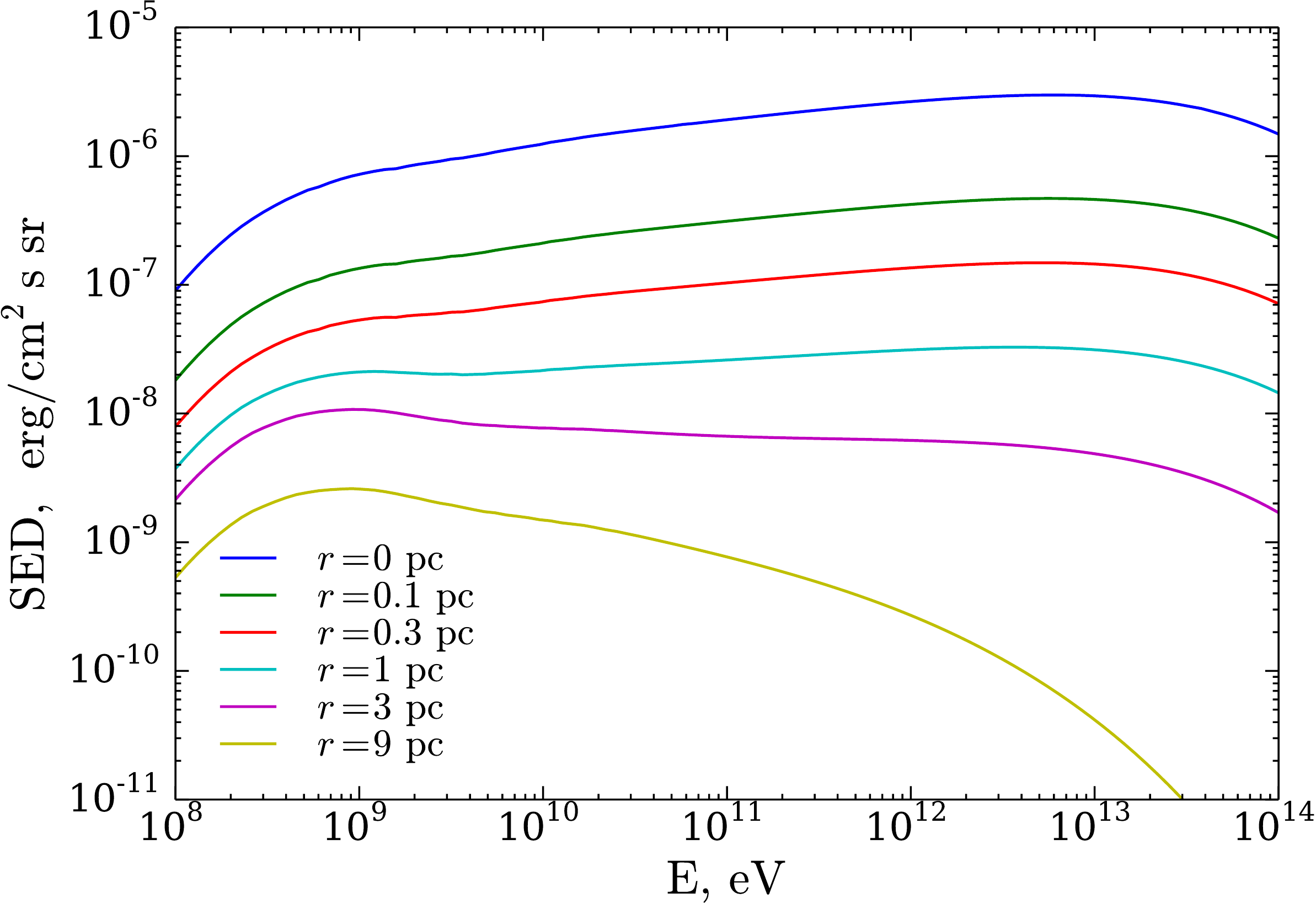}}
\caption{\label{fig:HmS} Energy spectra of gamma rays at different distances from 
the cosmic-ray source in the case of homogeneous cloud. The results are given for the  scenarios with  
slow  (left panel) and fast (right panel) proton diffusion.}
\end{center}
\end{figure*}

\begin{figure*}
\begin{center}
\includegraphics[bb=0 0 1440 504,width=1\textwidth]{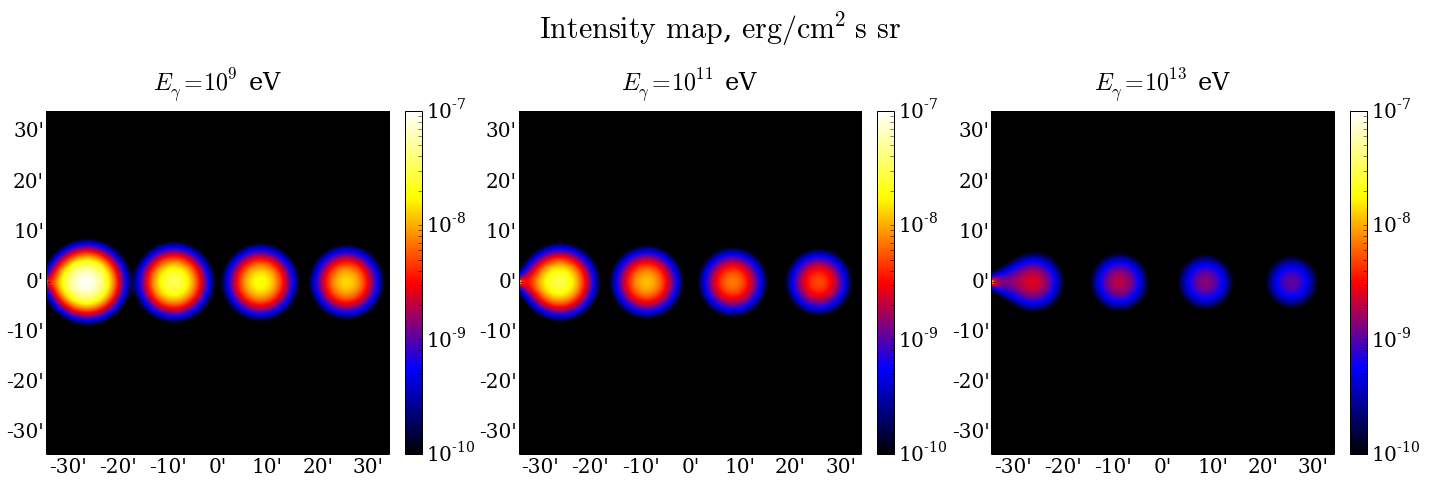}
\caption{\label{fig:CwLdIm} The intensity maps of gamma-ray emission from the group of clouds (without background) at various energies for the case of low diffusion coefficient. The cosmic-ray source is located in the centre of the left side.}
\includegraphics[bb=0 0 1440 504,width=1\textwidth]{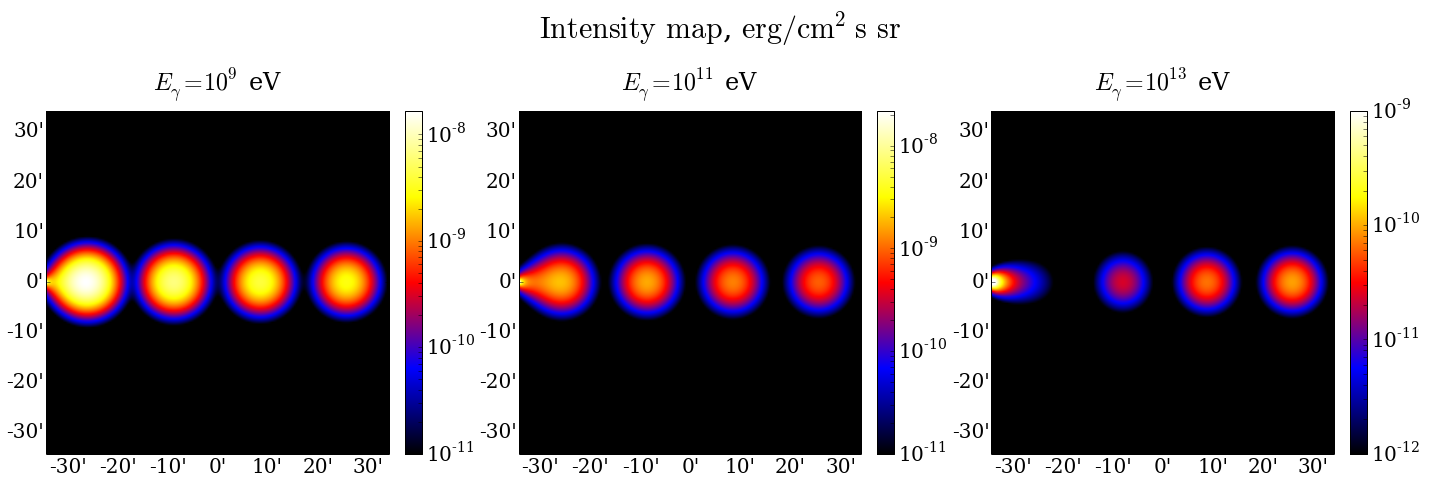}
\caption{\label{fig:CwHdIm} The same as in Fig.~\ref{fig:CwLdIm} for the case of high diffusion coefficient.}
\end{center}
\end{figure*}

\begin{figure*}
\begin{center}
\mbox{\includegraphics[bb=0 0 720 504,width=0.5\textwidth]{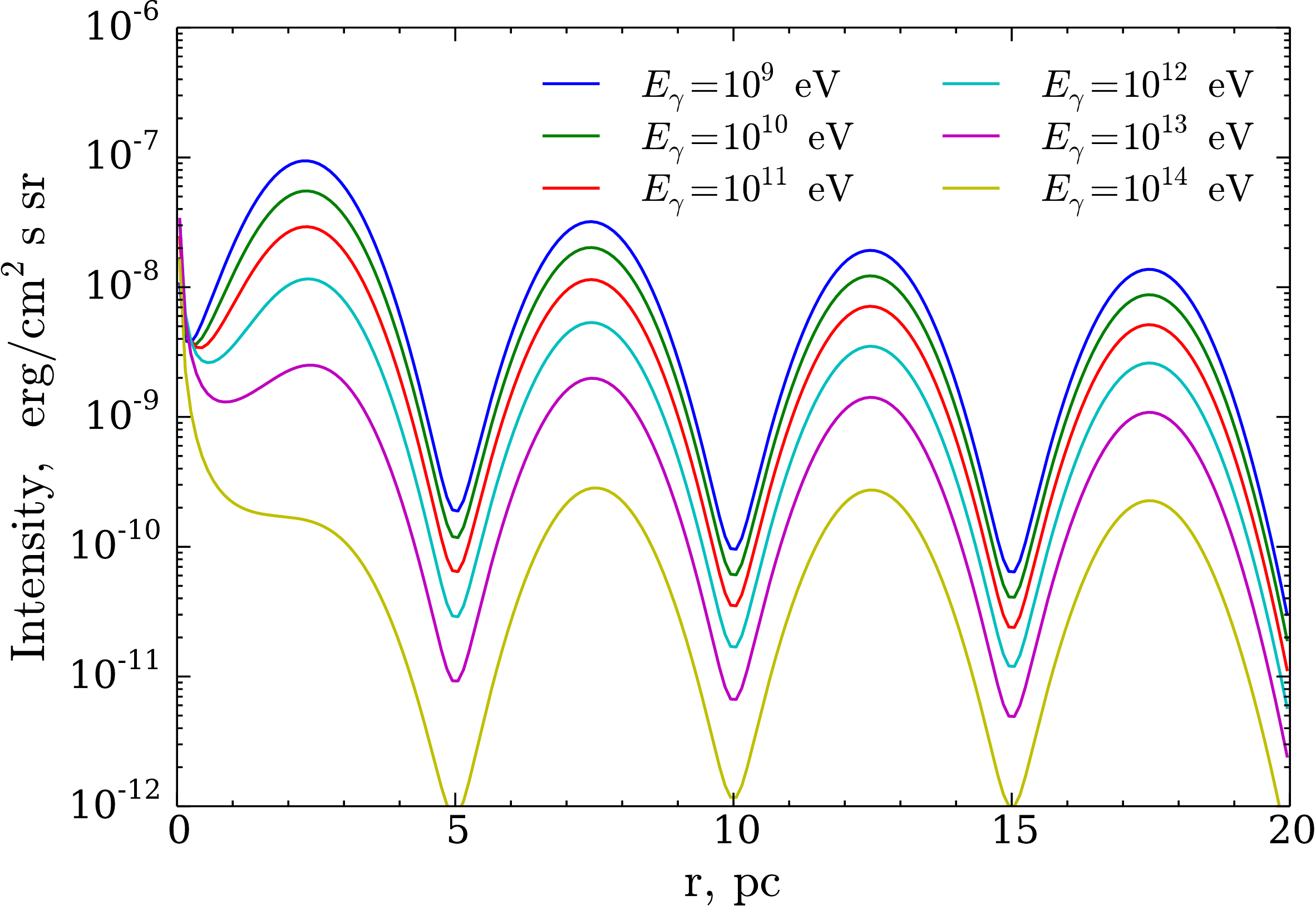}
\includegraphics[bb=0 0 720 504,width=0.5\textwidth]{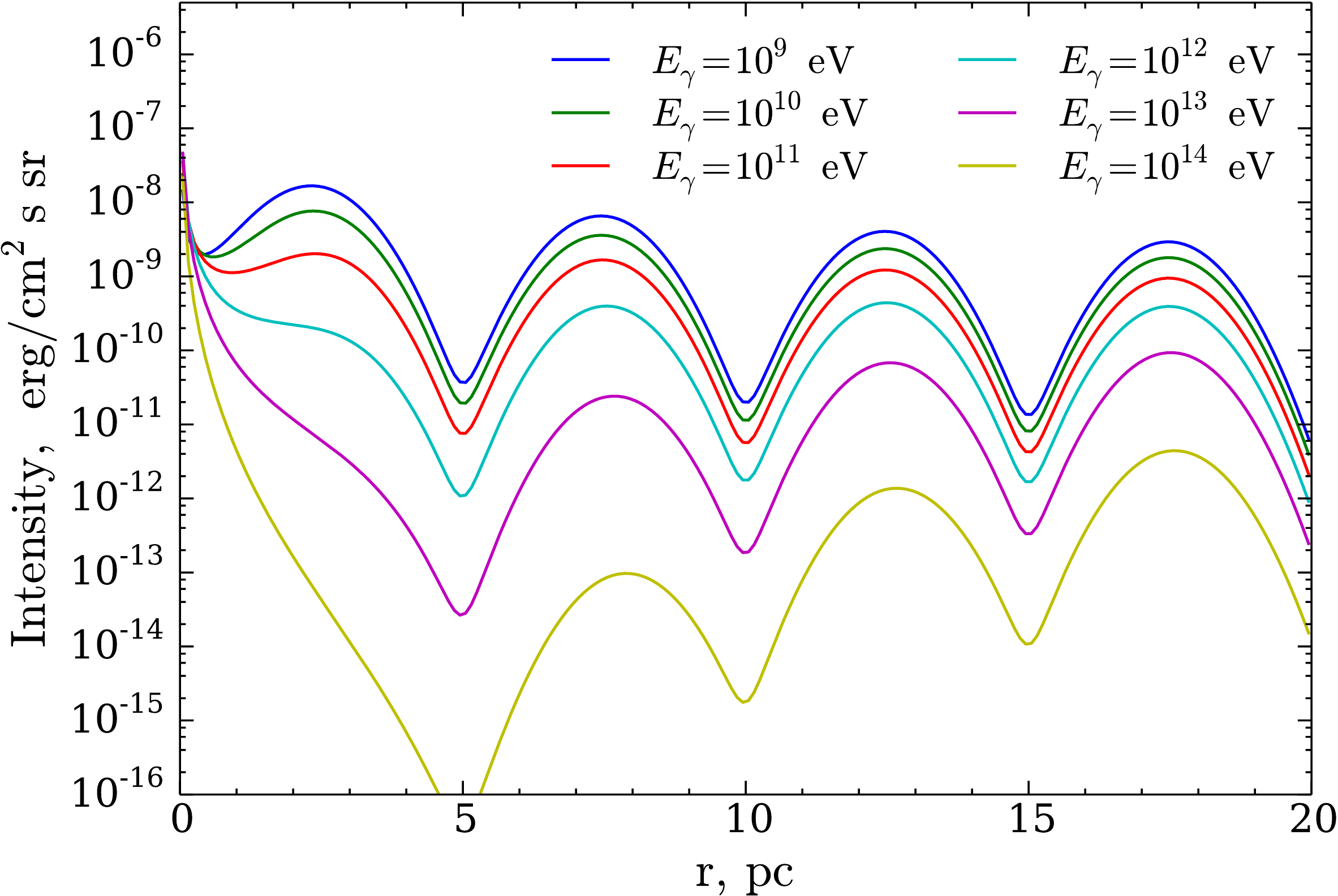}}
\caption{\label{fig:CwP} The radial intensity profiles for various energies in the case of group of clouds for the low (left panel) and high (right panel) diffusion coefficient. The profiles for energies $E=10^{9}$~eV, $E=10^{11}$~eV, and $E=10^{13}$~eV correspond to the intensity maps given in Fig.~\ref{fig:CwLdIm} for the left panel and Fig.~\ref{fig:CwHdIm} for the right panel.}
\end{center}
\end{figure*}

\begin{figure*}
\begin{center}
\includegraphics[bb=0 0 1440 504,width=1\textwidth]{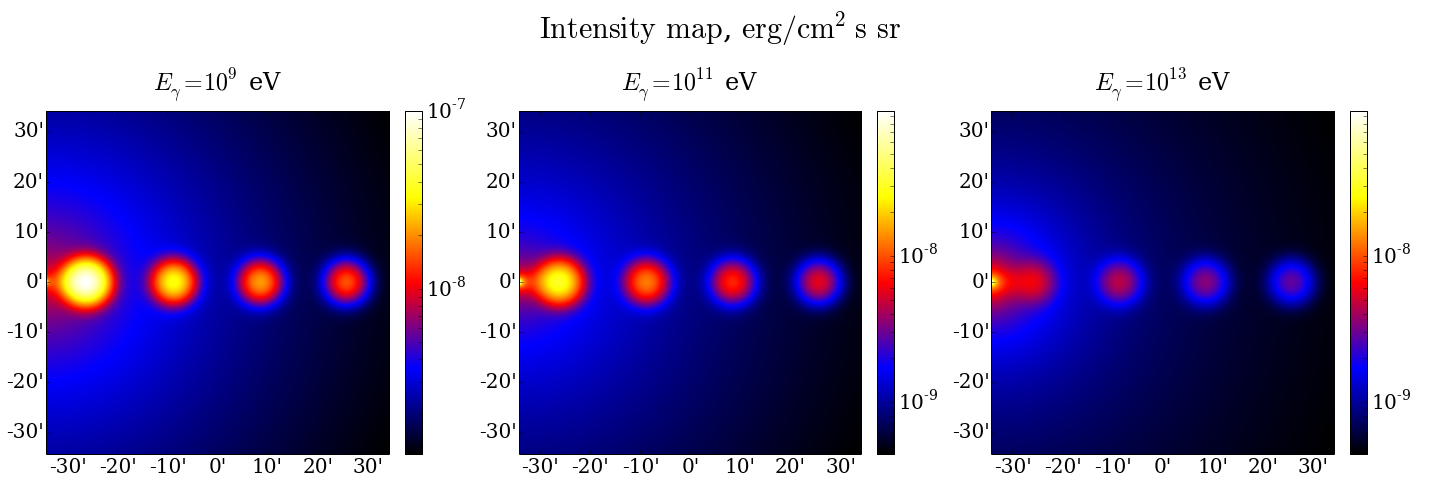}
\caption{\label{fig:CbLdIm}  The intensity maps of gamma-ray emission from the group of clouds and homogeneous background at various energies for the case of low diffusion coefficient. The cosmic-ray source is located in the centre of the left side.}
\includegraphics[bb=0 0 1440 504,width=1\textwidth]{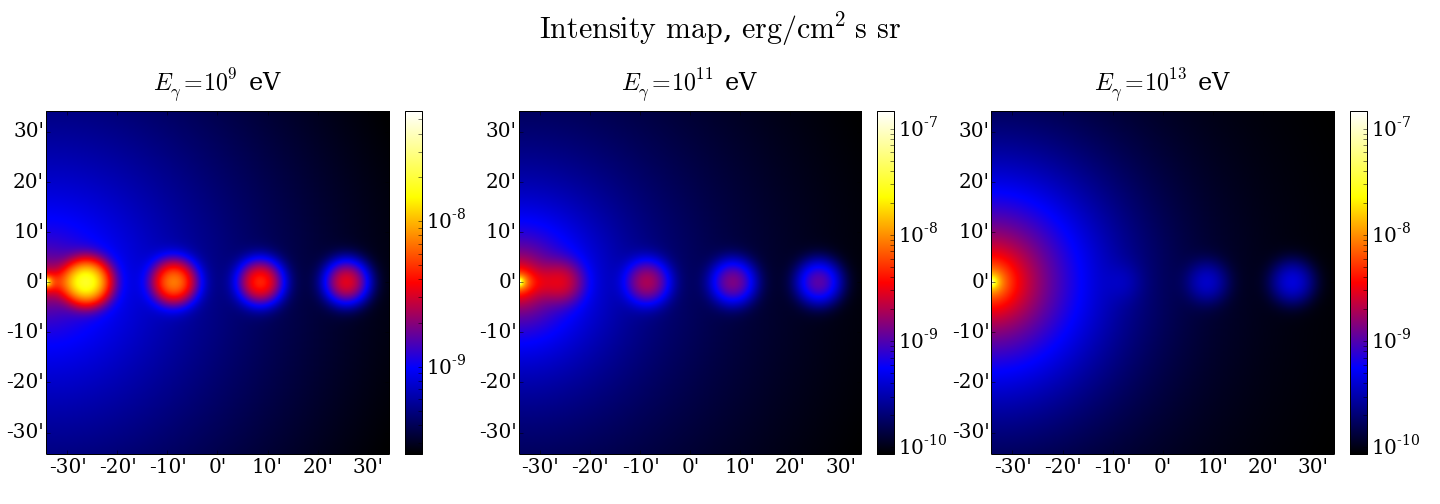}
\caption{\label{fig:CbHdIm} The same as in Fig.~\ref{fig:CbLdIm} for the case of high diffusion coefficient.}
\end{center}
\end{figure*}

\begin{figure*}
\begin{center}
\mbox{\includegraphics[bb=0 0 720 504,width=0.5\textwidth,angle=0]{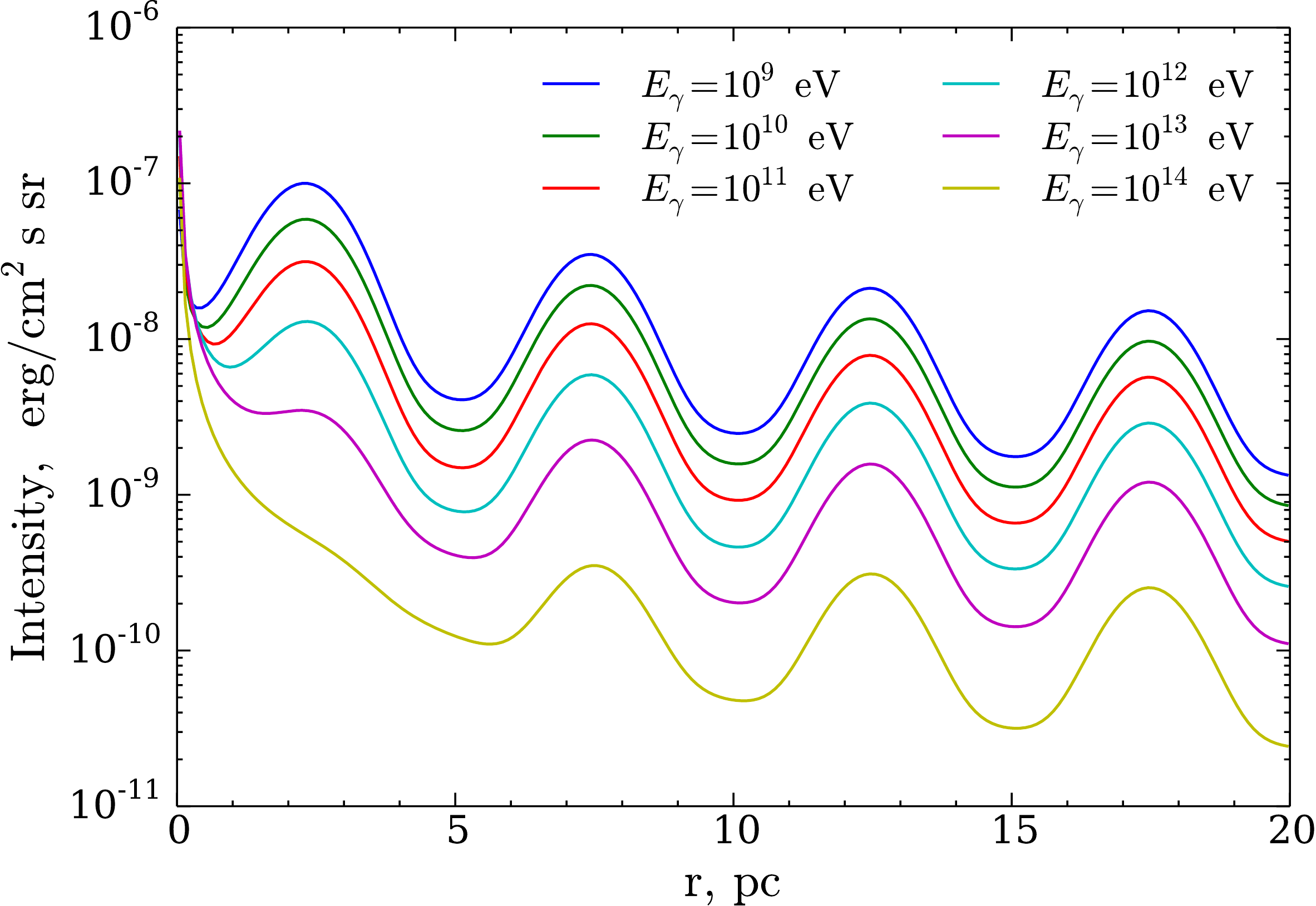}
\includegraphics[bb=0 0 720 504,width=0.5\textwidth,angle=0]{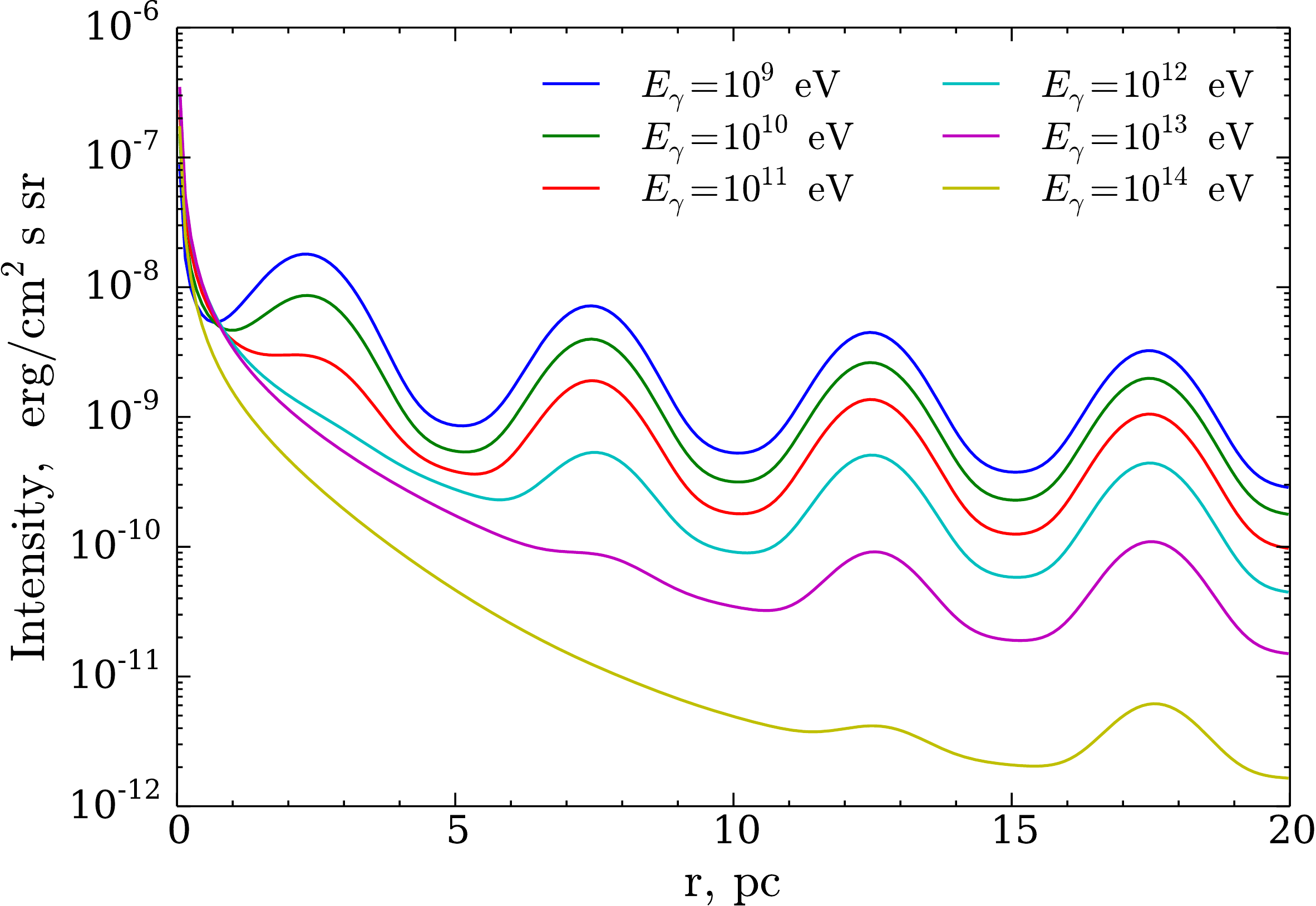}}
\caption{\label{fig:CbP} The radial intensity profiles for various energies in the case of group of clouds surrounded by homogeneous background for the low (left panel) and high (right panel) diffusion coefficient. The profiles for energies $E=10^{9}$~eV, $E=10^{11}$~eV, and $E=10^{13}$~eV correspond to the intensity maps given in Fig.~\ref{fig:CbLdIm} for the left panel and Fig.~\ref{fig:CbHdIm} for the right panel.}
\end{center}
\end{figure*}

\begin{figure*}
\begin{center}
\mbox{\includegraphics[bb=0 0 678 472,width=0.5\textwidth]{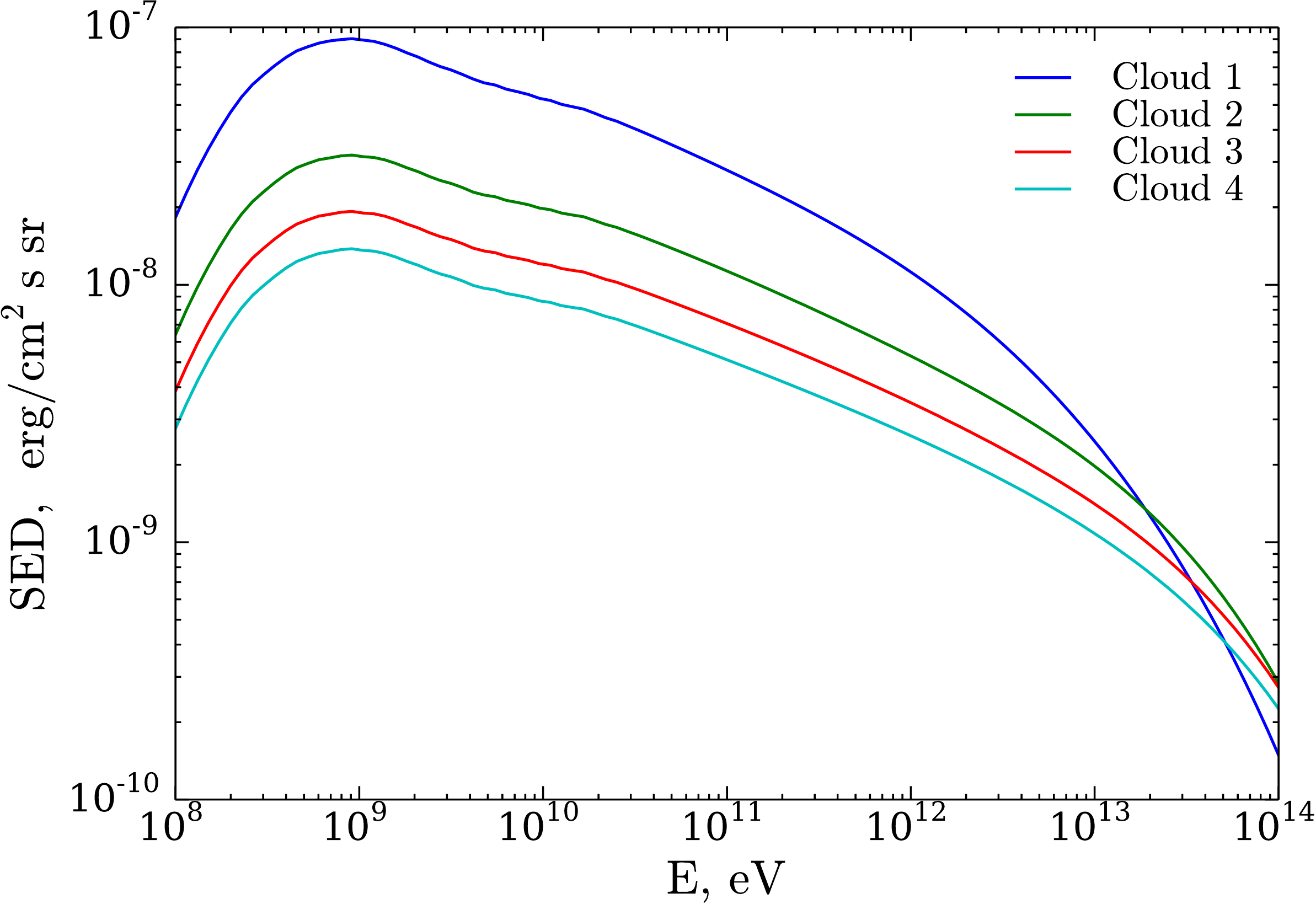}
\includegraphics[bb=0 0 678 472,width=0.5\textwidth]{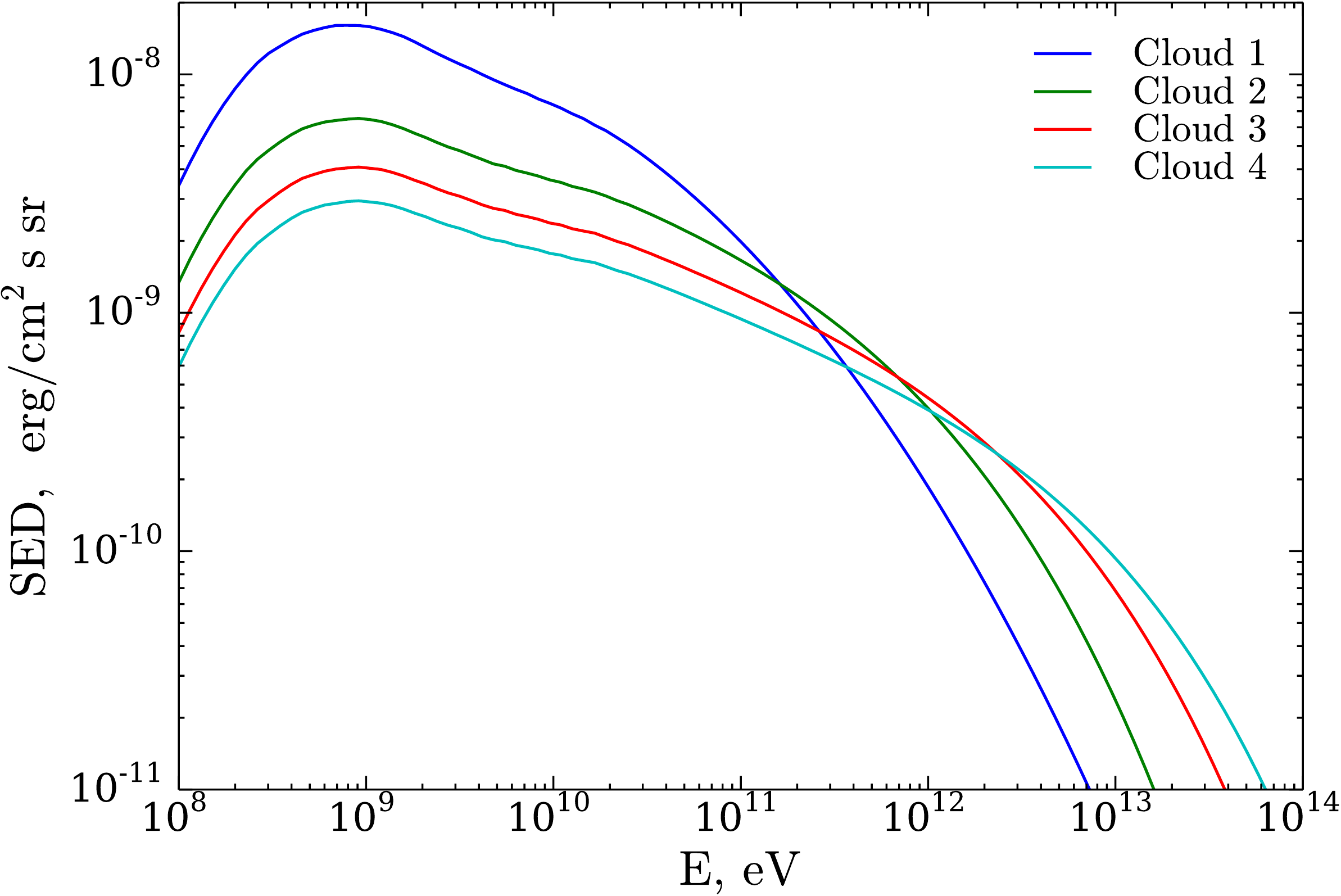}}
\caption{\label{fig:CwS} Energy spectra of gamma rays in the direction to the centres of the clouds in the case
of background absence for low (left panel) and high (right panel) diffusion coefficient. The clouds are numbered in the order of the distance from the source.}
\end{center}
\end{figure*}

\begin{figure*}
\begin{center}
\mbox{\includegraphics[bb=0 0 678 472,width=0.5\textwidth]{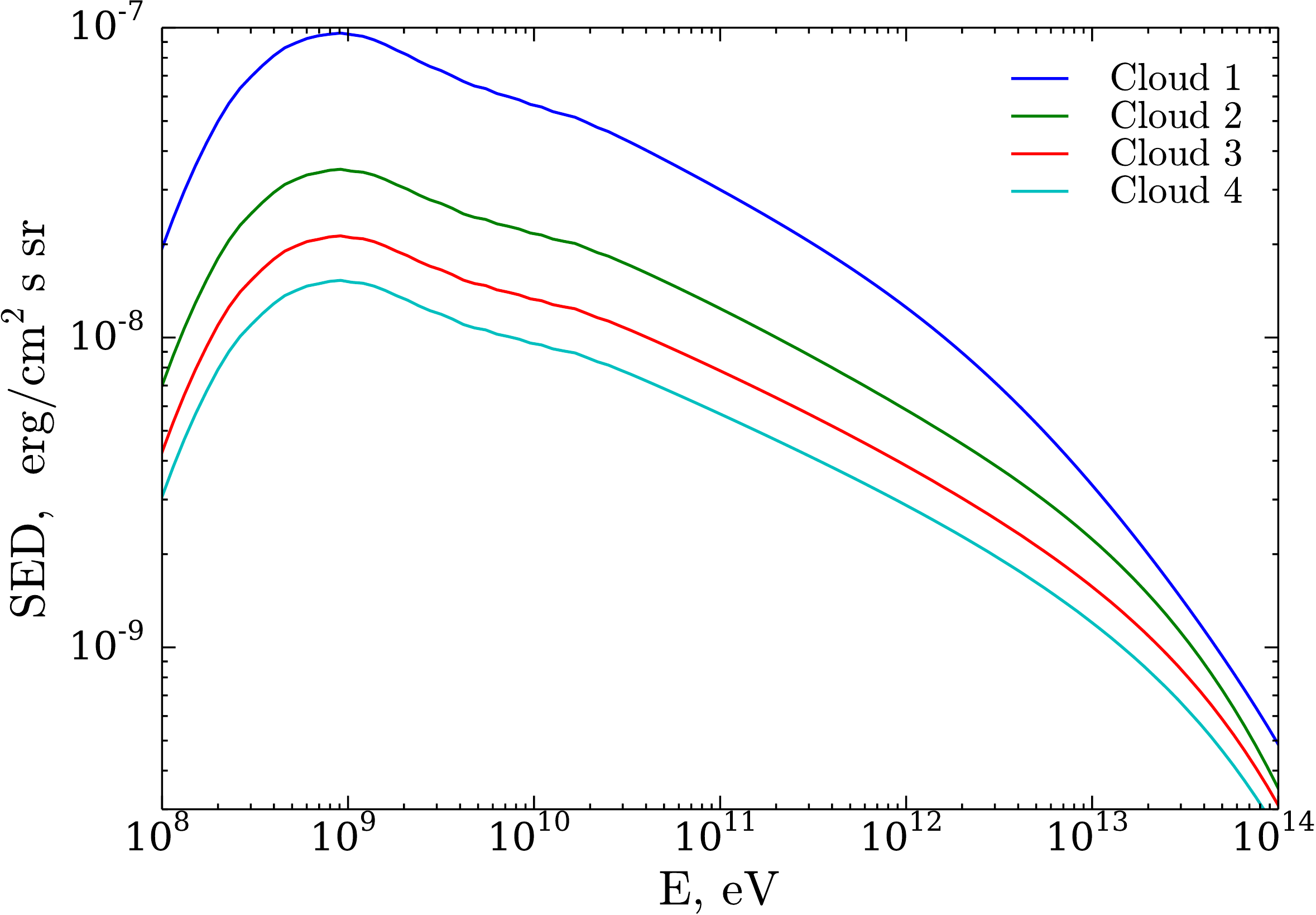}
\includegraphics[bb=0 0 678 472,width=0.5\textwidth]{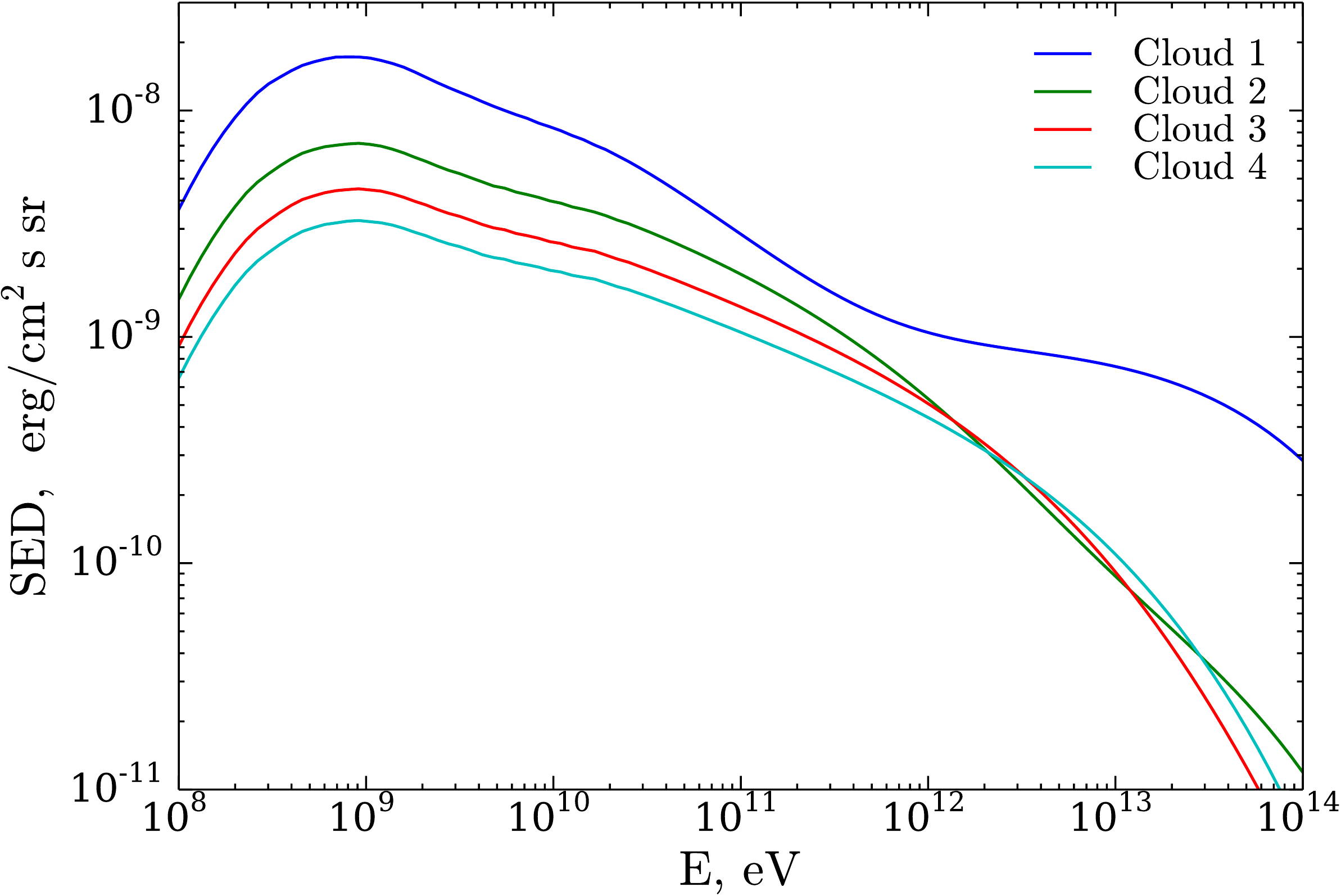}}
\caption{\label{fig:CbS} Energy spectra of gamma rays in the direction to the centres of the clouds in the case
of homogeneous background for low (left panel) and high (right panel) diffusion coefficient. The clouds are numbered in the order of the distance from the source.}
\end{center}
\end{figure*}


\begin{thebibliography}{15}
\expandafter\ifx\csname natexlab\endcsname\relax\def\natexlab#1{#1}\fi
\expandafter\ifx\csname bibnamefont\endcsname\relax
  \def\bibnamefont#1{#1}\fi
\expandafter\ifx\csname bibfnamefont\endcsname\relax
  \def\bibfnamefont#1{#1}\fi
\expandafter\ifx\csname citenamefont\endcsname\relax
  \def\citenamefont#1{#1}\fi
\expandafter\ifx\csname url\endcsname\relax
  \def\url#1{\texttt{#1}}\fi
\expandafter\ifx\csname urlprefix\endcsname\relax\def\urlprefix{URL }\fi
\providecommand{\bibinfo}[2]{#2}
\providecommand{\eprint}[2][]{\url{#2}}

\bibitem[{\citenamefont{{Dolginov} and {Toptygin}}(1967)}]{Dolginov1967}
\bibinfo{author}{\bibfnamefont{A.~Z.} \bibnamefont{{Dolginov}}}
  \bibnamefont{and}
  \bibinfo{author}{\bibfnamefont{I.}~\bibnamefont{{Toptygin}}},
  \bibinfo{journal}{Soviet Journal of Experimental and Theoretical Physics}
  \textbf{\bibinfo{volume}{24}}, \bibinfo{pages}{1195} (\bibinfo{year}{1967}).

\bibitem[{\citenamefont{{Aharonian} et~al.}(2010)\citenamefont{{Aharonian},
  {Kelner}, and {Prosekin}}}]{Aharonian2010}
\bibinfo{author}{\bibfnamefont{F.~A.} \bibnamefont{{Aharonian}}},
  \bibinfo{author}{\bibfnamefont{S.~R.} \bibnamefont{{Kelner}}},
  \bibnamefont{and} \bibinfo{author}{\bibfnamefont{A.~Y.}
  \bibnamefont{{Prosekin}}}, \bibinfo{journal}{\prd}
  \textbf{\bibinfo{volume}{82}}, \bibinfo{eid}{043002} (\bibinfo{year}{2010}),
  \eprint{1006.1045}.

\bibitem[{\citenamefont{{Syrovatskii}}(1959)}]{Syrovatskii1959}
\bibinfo{author}{\bibfnamefont{S.~I.} \bibnamefont{{Syrovatskii}}},
  \bibinfo{journal}{\sovast} \textbf{\bibinfo{volume}{3}}, \bibinfo{pages}{22}
  (\bibinfo{year}{1959}).

\bibitem[{\citenamefont{{Aloisio} et~al.}(2009)\citenamefont{{Aloisio},
  {Berezinsky}, and {Gazizov}}}]{Aloisio2009}
\bibinfo{author}{\bibfnamefont{R.}~\bibnamefont{{Aloisio}}},
  \bibinfo{author}{\bibfnamefont{V.}~\bibnamefont{{Berezinsky}}},
  \bibnamefont{and}
  \bibinfo{author}{\bibfnamefont{A.}~\bibnamefont{{Gazizov}}},
  \bibinfo{journal}{\apj} \textbf{\bibinfo{volume}{693}}, \bibinfo{pages}{1275}
  (\bibinfo{year}{2009}), \eprint{0805.1867}.

\bibitem[{\citenamefont{{Tautz}}(2013)}]{Tautz2013}
\bibinfo{author}{\bibfnamefont{R.~C.} \bibnamefont{{Tautz}}},
  \bibinfo{journal}{\aap} \textbf{\bibinfo{volume}{558}}, \bibinfo{eid}{A148}
  (\bibinfo{year}{2013}), \eprint{1309.7838}.

\bibitem[{\citenamefont{{Aloisio} and {Berezinsky}}(2004)}]{Aloisio2004}
\bibinfo{author}{\bibfnamefont{R.}~\bibnamefont{{Aloisio}}} \bibnamefont{and}
  \bibinfo{author}{\bibfnamefont{V.}~\bibnamefont{{Berezinsky}}},
  \bibinfo{journal}{\apj} \textbf{\bibinfo{volume}{612}}, \bibinfo{pages}{900}
  (\bibinfo{year}{2004}), \eprint{astro-ph/0403095}.

\bibitem[{\citenamefont{{Shalchi}}(2009)}]{Shalchi2009}
\bibinfo{editor}{\bibfnamefont{A.}~\bibnamefont{{Shalchi}}}, ed.,
  \emph{\bibinfo{title}{{Nonlinear Cosmic Ray Diffusion Theories}}}, vol.
  \bibinfo{volume}{362} of \emph{\bibinfo{series}{Astrophysics and Space
  Science Library}} (\bibinfo{year}{2009}).

\bibitem[{\citenamefont{{Blandford} and {Eichler}}(1987)}]{Blandford1987}
\bibinfo{author}{\bibfnamefont{R.}~\bibnamefont{{Blandford}}} \bibnamefont{and}
  \bibinfo{author}{\bibfnamefont{D.}~\bibnamefont{{Eichler}}},
  \bibinfo{journal}{\physrep} \textbf{\bibinfo{volume}{154}},
  \bibinfo{pages}{1} (\bibinfo{year}{1987}).

\bibitem[{\citenamefont{{Globus} et~al.}(2008)\citenamefont{{Globus}, {Allard},
  and {Parizot}}}]{Globus2008}
\bibinfo{author}{\bibfnamefont{N.}~\bibnamefont{{Globus}}},
  \bibinfo{author}{\bibfnamefont{D.}~\bibnamefont{{Allard}}}, \bibnamefont{and}
  \bibinfo{author}{\bibfnamefont{E.}~\bibnamefont{{Parizot}}},
  \bibinfo{journal}{\aap} \textbf{\bibinfo{volume}{479}}, \bibinfo{pages}{97}
  (\bibinfo{year}{2008}), \eprint{0709.1541}.

\bibitem[{\citenamefont{{Gabici} et~al.}(2009)\citenamefont{{Gabici},
  {Aharonian}, and {Casanova}}}]{Gabici2009}
\bibinfo{author}{\bibfnamefont{S.}~\bibnamefont{{Gabici}}},
  \bibinfo{author}{\bibfnamefont{F.~A.} \bibnamefont{{Aharonian}}},
  \bibnamefont{and}
  \bibinfo{author}{\bibfnamefont{S.}~\bibnamefont{{Casanova}}},
  \bibinfo{journal}{\mnras} \textbf{\bibinfo{volume}{396}},
  \bibinfo{pages}{1629} (\bibinfo{year}{2009}), \eprint{0901.4549}.

\bibitem[{\citenamefont{{Aharonian} and {Atoyan}}(1996)}]{Aharonian1996}
\bibinfo{author}{\bibfnamefont{F.~A.} \bibnamefont{{Aharonian}}}
  \bibnamefont{and} \bibinfo{author}{\bibfnamefont{A.~M.}
  \bibnamefont{{Atoyan}}}, \bibinfo{journal}{\aap}
  \textbf{\bibinfo{volume}{309}}, \bibinfo{pages}{917} (\bibinfo{year}{1996}).

\bibitem[{\citenamefont{{Nava} and {Gabici}}(2013)}]{Nava2013}
\bibinfo{author}{\bibfnamefont{L.}~\bibnamefont{{Nava}}} \bibnamefont{and}
  \bibinfo{author}{\bibfnamefont{S.}~\bibnamefont{{Gabici}}},
  \bibinfo{journal}{\mnras} \textbf{\bibinfo{volume}{429}},
  \bibinfo{pages}{1643} (\bibinfo{year}{2013}), \eprint{1211.1668}.

\bibitem[{\citenamefont{{Ptuskin}}(2006)}]{Ptuskin2006}
\bibinfo{author}{\bibfnamefont{V.}~\bibnamefont{{Ptuskin}}},
  \bibinfo{journal}{Journal of Physics Conference Series}
  \textbf{\bibinfo{volume}{47}}, \bibinfo{pages}{113} (\bibinfo{year}{2006}).

\bibitem[{\citenamefont{{Kafexhiu} et~al.}(2014)\citenamefont{{Kafexhiu},
  {Aharonian}, {Taylor}, and {Vila}}}]{Kafexhiu2014}
\bibinfo{author}{\bibfnamefont{E.}~\bibnamefont{{Kafexhiu}}},
  \bibinfo{author}{\bibfnamefont{F.}~\bibnamefont{{Aharonian}}},
  \bibinfo{author}{\bibfnamefont{A.~M.} \bibnamefont{{Taylor}}},
  \bibnamefont{and} \bibinfo{author}{\bibfnamefont{G.~S.}
  \bibnamefont{{Vila}}}, \bibinfo{journal}{\prd} \textbf{\bibinfo{volume}{90}},
  \bibinfo{eid}{123014} (\bibinfo{year}{2014}), \eprint{1406.7369}.

\bibitem[{\citenamefont{{Kelner} et~al.}(2006)\citenamefont{{Kelner},
  {Aharonian}, and {Bugayov}}}]{Kelner2006}
\bibinfo{author}{\bibfnamefont{S.~R.} \bibnamefont{{Kelner}}},
  \bibinfo{author}{\bibfnamefont{F.~A.} \bibnamefont{{Aharonian}}},
  \bibnamefont{and} \bibinfo{author}{\bibfnamefont{V.~V.}
  \bibnamefont{{Bugayov}}}, \bibinfo{journal}{\prd}
  \textbf{\bibinfo{volume}{74}}, \bibinfo{eid}{034018} (\bibinfo{year}{2006}),
  \eprint{astro-ph/0606058}.
  
\end{thebibliography}
\end{document}